\documentclass[longauth]{aa} 
\usepackage{graphicx}
\usepackage{txfonts}
\usepackage{lscape}    
\usepackage{color}
\usepackage{ulem}
\usepackage[dvipsnames]{xcolor}

\begin{document}

   \title{The Gaia-ESO survey: Mixing processes in low-mass stars traced by lithium abundance in cluster and field stars\thanks{Based on observations collected with the FLAMES instrument at
VLT/UT2 telescope (Paranal Observatory, ESO, Chile), for the Gaia-
ESO Large Public Spectroscopic Survey (188.B-3002, 193.B-0936, 197.B-1074).}\footnote{Tables 2, A.2 and A.3' are only available in electronic form
at the CDS via anonymous ftp to cdsarc.u-strasbg.fr (130.79.128.5)
or via http://cdsweb.u-strasbg.fr/cgi-bin/qcat?J/A+A/}}
\titlerunning{Li abundance and mixing in giant stars} 
\authorrunning{Magrini et al.}
\author{L. Magrini \inst{1},
           N. Lagarde\inst{2},
          C. Charbonnel\inst{3,4}, 
          E. Franciosini\inst{1}, 
          S. Randich\inst{1}, 
          R. Smiljanic\inst{5}, 
          G. Casali\inst{1,6},
          C. Viscasillas Vázquez\inst{7},
          L. Spina\inst{8}, 
          K. Biazzo\inst{9}, 
          L. Pasquini\inst{10},
          A. Bragaglia\inst{11}, 
          M. Van der Swaelmen\inst{1},
          G. Tautvai{\v s}ien{\. e}\inst{7},
          L. Inno\inst{12},
          N. Sanna\inst{1}, 
          L. Prisinzano\inst{13}, 
          S. Degl'Innocenti\inst{14,15}, 
          P. Prada Moroni\inst{14, 15}, 
          V. Roccatagliata\inst{14, 15, 1}, 
          E. Tognelli\inst{14,15}, 
          L. Monaco\inst{16},
          P. de Laverny\inst{17}, 
          E. Delgado-Mena\inst{18},
          M. Baratella\inst{19,8}, 
          V. D'Orazi\inst{8},
          A. Vallenari\inst{8}, 
          A. Gonneau\inst{20}, 
          C. Worley\inst{20},
          F. Jim\'enez-Esteban\inst{21},
          P. Jofre\inst{22},
          T. Bensby\inst{23}, 
          P. Fran\c cois\inst{24}, 
          G. Guiglion\inst{25}, 
          A. Bayo\inst{26,27}, 
          R.~D. Jeffries\inst{28}, 
          A.~S. Binks\inst{28}, 
          A.~Korn\inst{23}, 
          G. Gilmore\inst{20},
          F. Damiani\inst{13}, 
          E. Pancino\inst{1,29}, 
          G.~G. Sacco\inst{1},
          A. Hourihane\inst{20}, 
          L. Morbidelli\inst{1},
          S. Zaggia\inst{8}
          }

\institute{INAF-Osservatorio Astrofisico di Arcetri, Largo E.Fermi, 5. 50125, Firenze, Italy \email{laura.magrini@inaf.it} \and 
 Institut UTINAM, CNRS UMR 6213, Univ. Bourgogne Franche-Comt\'e, OSU THETA Franche-Comt\'e-Bourgogne, Observatoire
de Besan\c con, BP 1615, 25010, Besan\c con Cedex, France \and 
 Department of Astronomy, University of Geneva, Chemin de P\'egase 51, 1290 Versoix, Switzerland\and 
 IRAP, UMR 5277 CNRS and Universit\'e de Toulouse, 14, Av. E.Belin, 31400 Toulouse, France \and 
 Nicolaus Copernicus Astronomical Center, Polish Academy of Sciences, ul. Bartycka 18, 00-716, Warsaw, Poland\and 
 Dipartimento di Fisica e Astronomia, Università degli Studi di Firenze, via G. Sansone 1, 50019 Sesto Fiorentino (Firenze), Italy\and 
 Institute of Theoretical Physics and Astronomy, Vilnius University, Sauletekio av. 3, 10257 Vilnius, Lithuania\and 
 INAF - Padova Observatory, Vicolo dell'Osservatorio 5, 35122 Padova, Italy\and 
 INAF - Rome Observatory, Via Frascati, 33, Monte Porzio Catone (RM), Italy\and 
 ESO, Karl Schwarzschild Strasse 2, 85748 Garching, Germany\and 
 INAF - Osservatorio di Astrofisica e Scienza dello Spazio di Bologna, via Gobetti 93/3, 40129, Bologna, Italy\and 
 Science and Technology Department, Parthenope University of Naples, Centro Direzionale, Isola C4, 80143 Naples, Italy, INAF-Osservatorio Astronomico di Capodimonte, Salita Moraliello, Napoli\and 
 INAF - Osservatorio Astronomico di Palermo, Piazza del Parlamento 1, 90134, Palermo, Italy\and 
 Dipartimento di Fisica ‘E.Fermi’, Universitá di Pisa, Largo Bruno Pontecorvo 3, I-56127 Pisa, Italy\and 
 INFN, Sezione di Pisa, Largo Bruno Pontecorvo 3, I-56127 Pisa, Italy\and 
 Departamento de Ciencias Fisicas, Universidad Andres Bello, Fernandez Concha 700, Las Condes, Santiago, Chile \and 
 Universit\'e C\^{o}te d'Azur, Observatoire de la C\^{o}te d'Azur, CNRS, Laboratoire Lagrange, Nice, France\and 
 Instituto de Astrof\'isica e Ci\^encias do Espa\c{c}o, Universidade do Porto, CAUP, Rua das Estrelas, 4150-762 Porto, Portugal\and 
 Dipartimento di Fisica e Astronomia {\it Galileo Galilei}, Vicolo Osservatorio 3, I-35122, Padova, Italy \and 
 Institute of Astronomy, University of Cambridge, Madingley Road, Cambridge CB3 0HA, United Kingdom\and 
 Departamento de Astrof\'{\i}sica, Centro de Astrobiolog\'{\i}a (CSIC-INTA), ESAC Campus, Camino Bajo del Castillo s/n, E-28692 Villanueva de la Ca\~nada, Madrid, Spain\and 
 N\'ucleo de Astronom\'{i}a, Facultad de Ingenier\'{i}a y Ciencias, Universidad Diego Portales, Av. Ej\'ercito 441, Santiago, Chile\and 
 Lund Observatory, Department of Astronomy and Theoretical Physics, Box 43, SE-221 00 Lund, Sweden\and 
 GEPI, Observatoire de Paris, CNRS, Universit\'e Paris Diderot, 5 Place Jules Janssen, 92190 Meudon, France\and 
 Leibniz-Institut für Astrophysik Potsdam (AIP) An der Sternwarte 16, 14482 Potsdam\and 
 Instituto de Física y Astronomía, Facultad de Ciencias, Universidad de Valparaíso, Av. Gran Bretaña 1111, Valparaíso, Chile\and 
 Núcleo Milenio Formación Planetaria - NPF, Universidad de Valparaíso, Av. Gran Bretaña 1111, Valparaíso, Chile\and 
 Astrophysics Group, Keele University, Keele, Staffordshire ST5 5BG, United Kingdom \and 
 Space Science Data Center - Agenzia Spaziale Italiana, via del Politecnico, s.n.c., I-00133, Roma, Italy} 
   \date{}

 
  \abstract
   {}
   {We aim to  constrain  the mixing processes in low-mass stars by investigating the behaviour of the Li surface abundance after the main sequence. We take advantage of the data from the sixth internal data release of Gaia-ESO, {\sc idr6}, and from the {\it Gaia}  Early Data Release 3, {\sc edr3}.  }
   {We select a sample of main sequence, sub-giant, and  giant stars in which Li abundance is measured by the Gaia-ESO survey, belonging to 57 open clusters with ages from 120~Myr to about 7 Gyr and to Milky Way fields, covering a range in [Fe/H] between $\sim\,-1.0$ and $\sim\,+0.5$~dex. 
   We study the behaviour of the Li abundances as a function of stellar parameters. We infer the masses of giant stars in clusters from the main-sequence turn-off masses, and for field stars through comparison with stellar evolution models using a maximum-likelihood technique.  
   We compare the observed Li behaviour in field giant stars and in giant stars belonging to individual clusters with the predictions of a set of classical models and of models with mixing induced by rotation and thermohaline instability.   
   }
   {The comparison with stellar evolution models confirms that classical models cannot reproduce the lithium abundances observed in the metallicity and mass regimes covered by the data. 
   The models that include  the effects of both rotation-induced mixing and thermohaline instability account for the Li abundance trends observed in our sample, in all metallicity and mass ranges.  The differences between the results of the classical models and of the rotation models largely differ (up to ~2 dex), making  lithium the best element to constrain stellar mixing processes  in low-mass stars. We discuss 
   the nature of a sample of Li-rich stars.  
     }
   {We demonstrate that the evolution of the surface abundance of Li in giant stars is a powerful tool to constrain theoretical stellar evolution models, allowing us to distinguish the impact of different mixing processes. For stars with well-determined masses, we find a better agreement between observed surface abundances and models with rotation-induced and thermohaline mixings, the former dominating during the main sequence and the first phases of the post-main sequence evolution and the latter after the bump in the luminosity function.    }


   \maketitle
%

\section{Introduction}
\label{section:introduction}

Big Bang nucleosynthesis produced mostly H and He, together with a small amount of the lithium-7 isotope  \citep[hereafter Li; e.g.][]{2004ApJ...600..544C,GP13,2013AIPC.1560..314O, pitrou18}. However, the Li that one observes in the present-time Universe is only in part the one originally produced during the Big Bang, as its abundance is modified by a number of constructive and destructive processes which make Li one of the elements with the most complex history \citep[e.g.][and references therein]{Matteucci1995,Romano2001,Travaglio2001,2012A&A...542A..67P, bensby18, 2019MNRAS.489.3539G, randich20, 2020MmSAI..91..142S, 2021FrASS...8....6R}. 

One of the difficulties in tracing the history of cosmic Li is that, with the exception of the early pre-main sequence (PMS) phases, stars rarely exhibit the original Li with which they formed. This fragile element is destroyed by proton captures in stellar interiors when the temperature is of the order of $\sim$2.5~10$^6$~K or higher. Depending on the mass and metallicity of the star, photospheric Li can be significantly depleted already on the pre-main sequence (PMS), during the proto-stellar accretion phase \citep{tognelli20} and along the Hayashi track,  and/or on the main sequence (MS), due to  several mechanisms that have the potential to transport the photospheric material into hotter layers where Li can be burned: atomic diffusion, overshooting, rotation-induced mixing, internal gravity waves  and other types of magneto-hydrodynamical instabilities that are not included in the so-called classical evolution models \citep{1986ApJ...302..650M,CharbonneauMichaud1990,1990ApJ...359L..55S,Richard1996,2000IAUS..198...61D, denissenkov03, 2010IAUS..268..365T,2012A&A...539A..70E,Castroetal2016,2016ApJ...829...32S,2017ApJ...845L...6B,2021A&A...646A.160D,2021A&A...646A..48D}.

After the MS, convection sinks inside the stars bringing material that has been partially nuclear-processed in the stellar interior to the surface. This enriches the external layers in $^{13}$C, $^{14}$N and $^{3,4}$He, and dilutes Li. According to the classical model by \citet{1967ApJ...147..624I}, during this so-called first dredge-up event (FDU), 
the surface Li abundance decreases by a factor from 30 to 60, depending on the stellar mass and metallicity.
Starting from the present interstellar medium abundance of A(Li)\footnote{A(Li)=log($\frac{X(Li)}{X(H)}\cdot \frac{A_{\rm H}}{A_{\rm Li}}$)+12, where X and A are the mass fraction and the atomic mass} = 3.3~dex \citep[the value found in Solar System meteorites and considered the reference limit for Population I dwarf stars; ][]{asplund09},   the Li abundance of red giant stars is thus expected to
decrease at least down to a value A(Li)$\sim$1.5~dex. 
Models that include some of the above-mentioned transport processes already acting on the MS predict lower post-FDU values of lithium abundances. These models better agree with the observations of subgiant and giant stars \citep[i.e., that Li depletion appears at hotter effective temperature and is larger than in classical models; see e.g.][]{brown89,1995MmSAI..66..387B,palacios03,mallik03,pasquini04,2006A&A...450.1173L,gonzalez09,2009AJ....138.1171A,c20}. 

Finally, classical models do not predict any decreasing trend of Li abundance in the subsequent evolutionary phases, 
although Li is observed to drop again after the luminosity bump on the red giant branch \citep[RGB; e.g.][]{1998A&A...332..204C,gratton00,lind09}. This is likely caused by the activation of thermohaline (double diffusive) instability, which could also affect carbon and nitrogen abundances \citep[e.g.][]{cczahn07,2009ApJ...696.1823D,CL10,lattanzio15, 2018ApJ...862..136S,2017MNRAS.469.4600H}. 

While some non-classical stellar models manage to reproduce the main Li trends described above, they still suffer from serious shortcomings. 
For example, they fail to simultaneously reproduce the internal rotation profiles of sub-giant and giant stars as depicted by asteroseismology \citep[e.g.][]{2013A&A...549A..74M,2013A&A...555A..54C,2017A&A...599A..18E,2019A&A...621A..66E}. In addition, different prescriptions for thermohaline mixing are required to explain the surface abundance of Li and C in low-metallicity bright red-giant stars \citep[e.g.][]{2015MNRAS.450.2423A,2017MNRAS.469.4600H}. 
The difficulty is that macroscopic magnetic hydrodynamic (MHD) transport processes act on a broad range of spatial and time scales which cannot be handled numerically when computing secular evolution \citep[e.g.][]{2013LNP...865...23M}. One-dimensional (1D) stellar models thus rely on simplified prescriptions for, e.g., turbulence and magneto-hydrodynamic instabilities that are, in the best case, "educated" from numerical and laboratory experiments which are however still far from reproducing stellar interior conditions \citep[e.g.][]{1999A&A...347..734R,palmerini11, 2013A&A...551L...3P,2015sf2a.conf..419P,2016ApJ...821...49G,2018A&A...620A..22M,Sengupta18,2021arXiv210308072G}. 

In this framework, observations of large samples of stars with available Li abundances provide fundamental constraints to models. 
However, most studies focused on field stars, including only small numbers of star clusters \citep[see. e.g.][]{lambert80, Balachandran90, pasquini04, lind09, smi09, canto11}. So far, a homogeneous analysis of Li in both field and cluster populations is missing, even in  large spectroscopic surveys, such as the GALactic Archaeology with HERMES survey \citep[GALAH][]{buder20}.  

In the present work, we take advantage of the Gaia-ESO database \citep{Gil, randich13} for the sixth internal data release ({\sc idr6}), which includes homogeneously-determined Li abundances in stars of open clusters and in the field. With these data, we investigate the Li abundance evolution after the MS over a large range of [Fe/H] and stellar masses. 
In particular, for clusters metallicity and age, and consequently the masses of their stars at the main sequence turn-off (MSTO) and RGB, can be estimated more accurately than for field stars, therefore they allow a more accurate comparison with the results of theoretical models. In addition, the observed star clusters are usually younger than field stars, allowing us to map higher mass ranges. 

The paper is structured as follows. 
In Sect.~\ref{section:abundance_and_sample} 
we present the abundance analysis and the sample selection. In Sect.~\ref{section:cataloguecomparison}, we compare the Gaia-ESO {\sc idr6} results with other catalogues. 
In Sect.~\ref{section:datamodelcomparison}, we study the behaviour of Li abundances along the RGB in field stars with  masses determined with a maximum-likelihood method using a large and homogeneous grid of stellar models, and in members of individual star clusters. We compare our data with model predictions and discuss the impact of rotation-induced mixing and thermohaline instability. 
In Sect.~\ref{sec:lithium:rich}, we identify Li-rich giants, discuss their properties and the effect of stellar rotation. 
Finally, in Sect.~\ref{section:SummaryConclusions} 
we give our summary and conclusions. 

\section{Abundance analysis and sample selection}
\label{section:abundance_and_sample}

\subsection{Li abundance determination}
\label{subsection:Liabundancedetermination}

We use data from {\sc idr6} of the Gaia-ESO Survey, derived from both the UVES spectra with resolving power R$=$47,000 and spectral range 480.0$-$680.0~nm, and the GIRAFFE HR15N spectra (R$\sim$19,000), covering the wavelength range 647$-$679~nm. Both types of spectra have been reduced and analysed by the Gaia-ESO consortium. The data reduction and analysis process have been described in several papers \citep[see, e.g.][]{sacco14, smi14, jackson15, lanzafame15};
we recall here the main steps. The pipelines for data reduction, as well as radial and rotational velocity determinations, are run 
at INAF-Arcetri for UVES \citep{sacco14} using the FLAMES-UVES ESO public pipeline, and 
at the Cambridge Astronomy Survey Unit (CASU)
for GIRAFFE.
The spectral analysis is shared among different working groups (WGs), to which spectra are assigned on the basis of the stellar-type, instrument, and setup. The data discussed in the present paper have been analysed by WG10, WG11, and WG12 which are in charge of the analysis of the UVES and GIRAFFE spectra of F-G-K (and M for WG12) stars both in the field of the Milky Way (MW) and in open clusters. The spectra in each WG are analysed with a strategy based on multiple pipelines, as described in Worley et al. (in preparation), \citet{smi14} and \citet{lanzafame15} for WG10, WG11 and WG12, respectively. 
Finally, the results from the different WGs are homogenised using a database of calibrators, e.g., benchmark stars and open/globular clusters selected following the calibration strategy by \citet{pancino17} and adopted for the homogenisation by WG15 (Hourihane et al. in preparation). 
The recommended parameters and abundances are distributed in the {\sc idr6} catalogue, which includes those used in the present work: atmospheric stellar parameters $T_\mathrm{eff}$, $\log g$, and/or $\gamma$, the surface gravity index  based on the ratios of high-gravity and low-gravity lines in the spectral region 675.0-678.0~nm\, and  defined in \citet{damiani14},  metallicity [Fe/H], lithium abundances (measurements or upper limits), radial velocities (RVs), and projected equatorial rotational velocities (v$sin$i).  

The lithium abundance is measured from the doublet lines at 670.8~nm. At the resolution of GIRAFFE, this line is blended with the nearby FeI line at 670.74~nm, but the two components can be well separated in UVES. In {\sc idr6}, the Li abundances from both UVES and Giraffe spectra were derived in a homogeneous way.
Lithium equivalent widths (EWs) were measured by gaussian 
fitting of the lithium doublet components and the FeI line,
and then converted into abundances using a set of {\it ad hoc} curves of growth (Franciosini et al., in prep.), specifically derived for the Gaia-ESO survey with a grid of synthetic spectra computed as in \citet{delaverny12} and \citet{Guiglion16} and based on the MARCS model atmospheres in the following ranges: $3000\le T_\mathrm{eff}\le 8000$~K, $0.5\le\log~g\le 5.0$, $-2.50\le\,$[Fe/H]$\,\le +0.50$ and $-1.0\le\,$A(Li)$\,\le 4.0$. 
In the case of GIRAFFE, where only the total blended Li$+$Fe EW can be measured, 
the Li-only EW was first computed by applying a correction for the Fe blend, measured on the same synthetic spectra used to derive the curves of growth.
When the line is not visible (or just barely visible), an upper limit to the EW, equal to the uncertainty, or to the measured EW if higher, is given.

\begin{figure}[h!]
\centering
\includegraphics[width=\hsize]{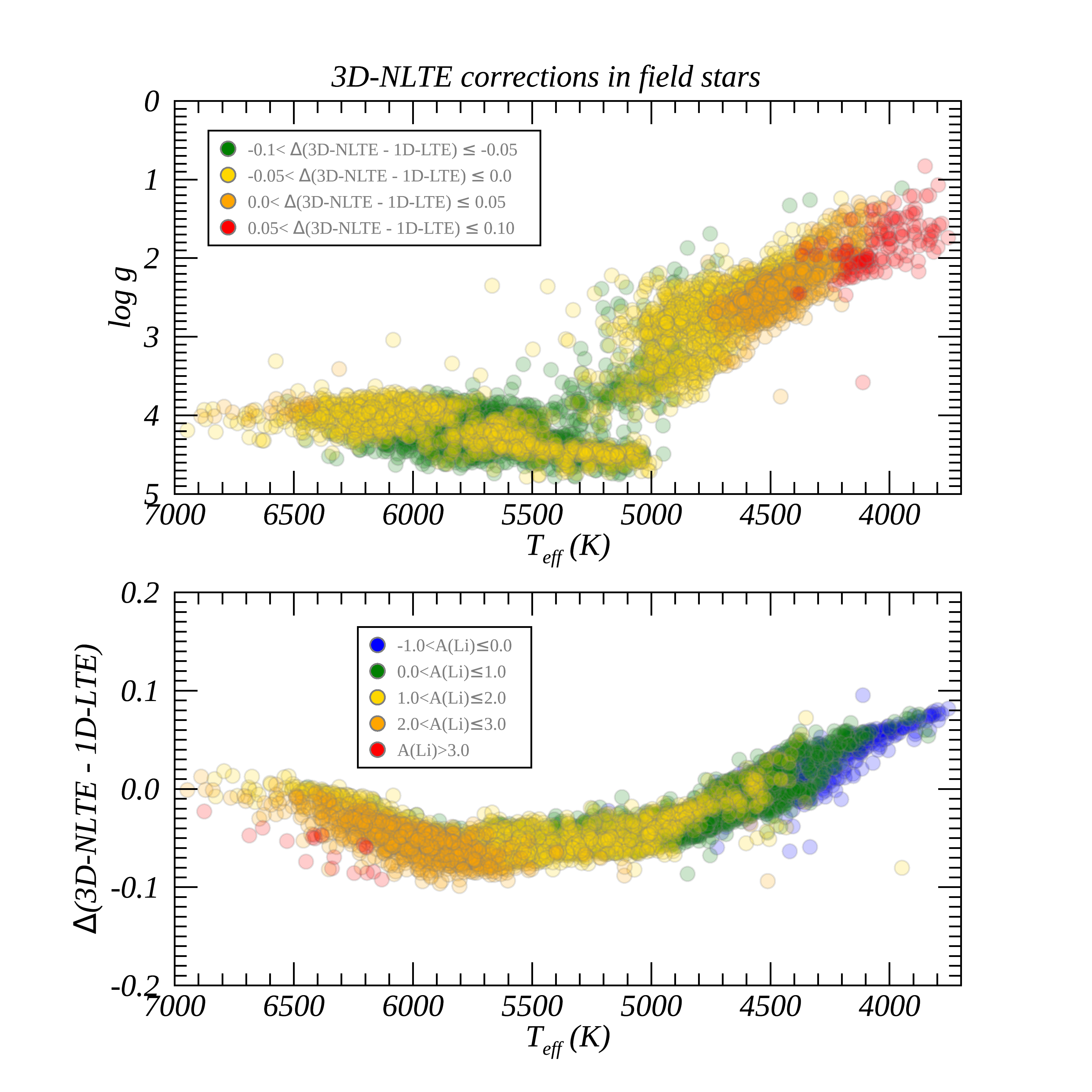}
\caption{Upper panel: Kiel diagram of the field stars, colour coded by $\Delta$(A(Li)$_{\rm 3D-NLTE}$-A(Li)$_{\rm 1D-LTE}$). Lower panel: $\Delta$(A(Li)$_{\rm 3D-NLTE}$-A(Li)$_{\rm 1D-LTE}$) versus $T_\mathrm{eff}$ for field stars colour-coded by A(Li).}
\label{fig:nlte:fields}
\end{figure}

\begin{figure}[h!]
\centering
\includegraphics[width=\hsize]{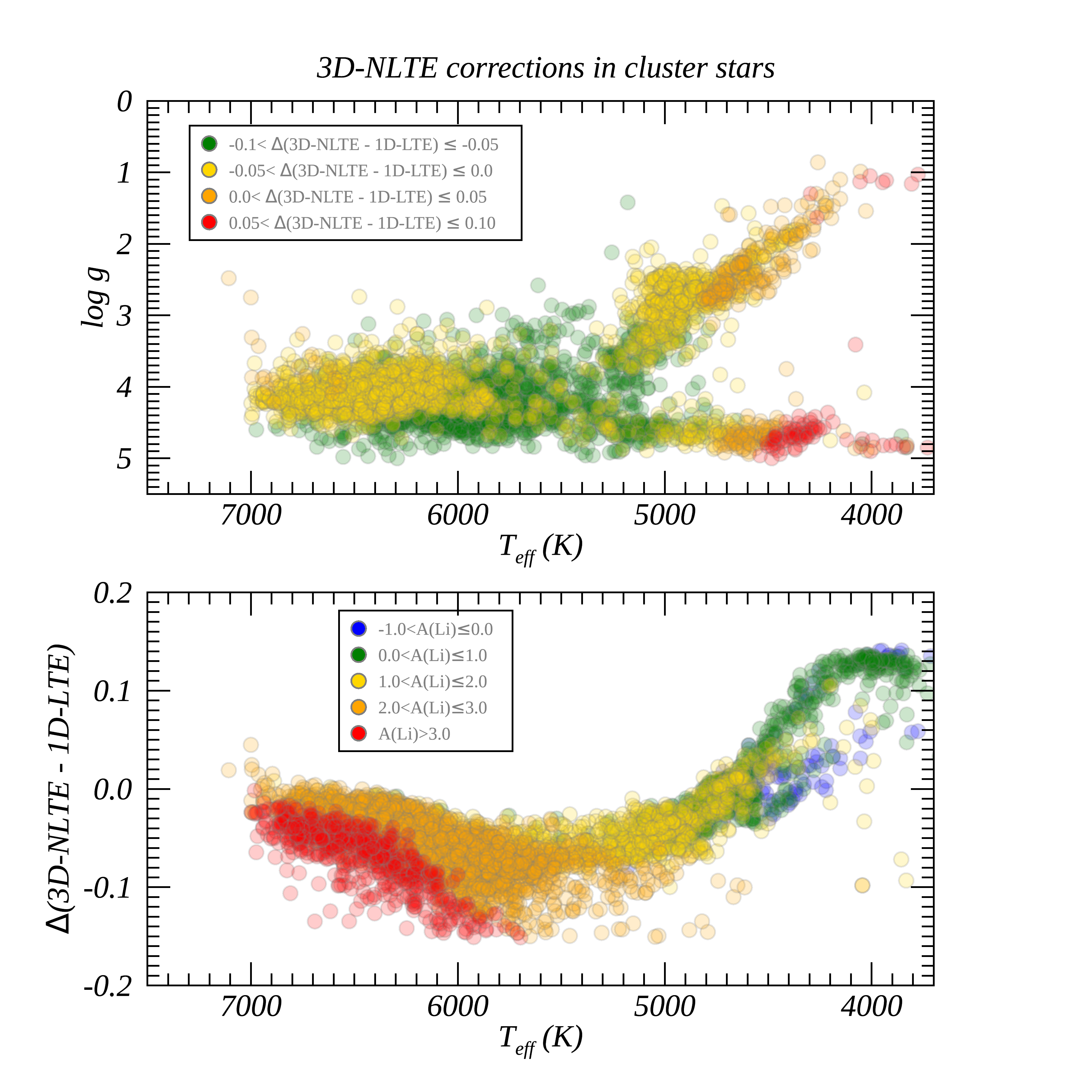}
\caption{Same as Fig.~\ref{fig:nlte:fields} but for stars in open clusters.}
\label{fig:nlte:clusters}
\end{figure}

The Gaia-ESO abundances are determined in the Local Thermodynamic Equilibrium (LTE) approximation. 
We estimate the typical effect on Li abundances introduced by the LTE approximation in 1D hydrostatic atmospheres following  
\citet{wang21}, who compute a 3D NLTE Li grid spanning the 
parameter range  for FGK-type dwarfs and giants. 
The 3D NLTE corrections can increase or decrease A(Li) by a few tenths of a dex in the typical ranges of parameters of our sample of giant stars. We compute them for the sample of stars for which all stellar parameters are available, using the code and grids provided by \citet{wang21}.  
In Figs.~\ref{fig:nlte:fields} and \ref{fig:nlte:clusters}, we show the effect of the correction $\Delta$(A(Li)$_{\rm 3D-NLTE}$-A(Li)$_{\rm 1D-LTE}$)) in the Kiel diagram, and as a function of $T_\mathrm{eff}$ for field stars and open clusters, respectively.  
For both samples the effect is within $\pm$0.1~dex, depending on $T_\mathrm{eff}$, and almost negligible for MSTO stars and for giant stars hotter than 4200~K. In the next sections, we adopt the 1D LTE Gaia-ESO Li abundances, available also for stars for which log~g has not been determined.  


\subsection{Sample selection}
\label{sec:sample}
The present work analyses the Gaia-ESO {\sc idr6}  sample of MS, sub-giant, and giant stars with available Li determination, focusing on the post-MS evolution of lithium surface abundances.   
The spectral ranges of the HR15N and U580 setups allow us to measure the Li doublet lines.  
In particular, the GIRAFFE setup HR15N is dedicated in Gaia-ESO to the study of stars in open clusters. 
However, since the target selection of member stars in clusters observed with GIRAFFE is unbiased and inclusive, many non-member contaminants, including in particular giant stars,  are present in the Gaia-ESO database. The contamination by distant field giants is even more important in the field of view of the youngest star clusters, due to the similar colours and hence position in the Colour-Magnitude diagrams that were used for target selection.  
We take advantage of this favourable configuration to build a large sample of high-resolution spectra of field and cluster stars with Li measurements in a wide range of metallicity. 
We broadly define giant stars as those with $T_\mathrm{eff}\le 5400$~K and $\log g\le 3.8$ (or $\gamma \ge 0.98$, if $\log g$ is not available), then the remaining sample includes sub-giants (with $5400 \le T_\mathrm{eff}\le 6000$~K and $\log g\le 4$, although the limits are difficult to determine precisely) and MS stars. 

\subsubsection{The open cluster sample}

\begin{figure}[h!]
\centering
\includegraphics[width=\hsize]{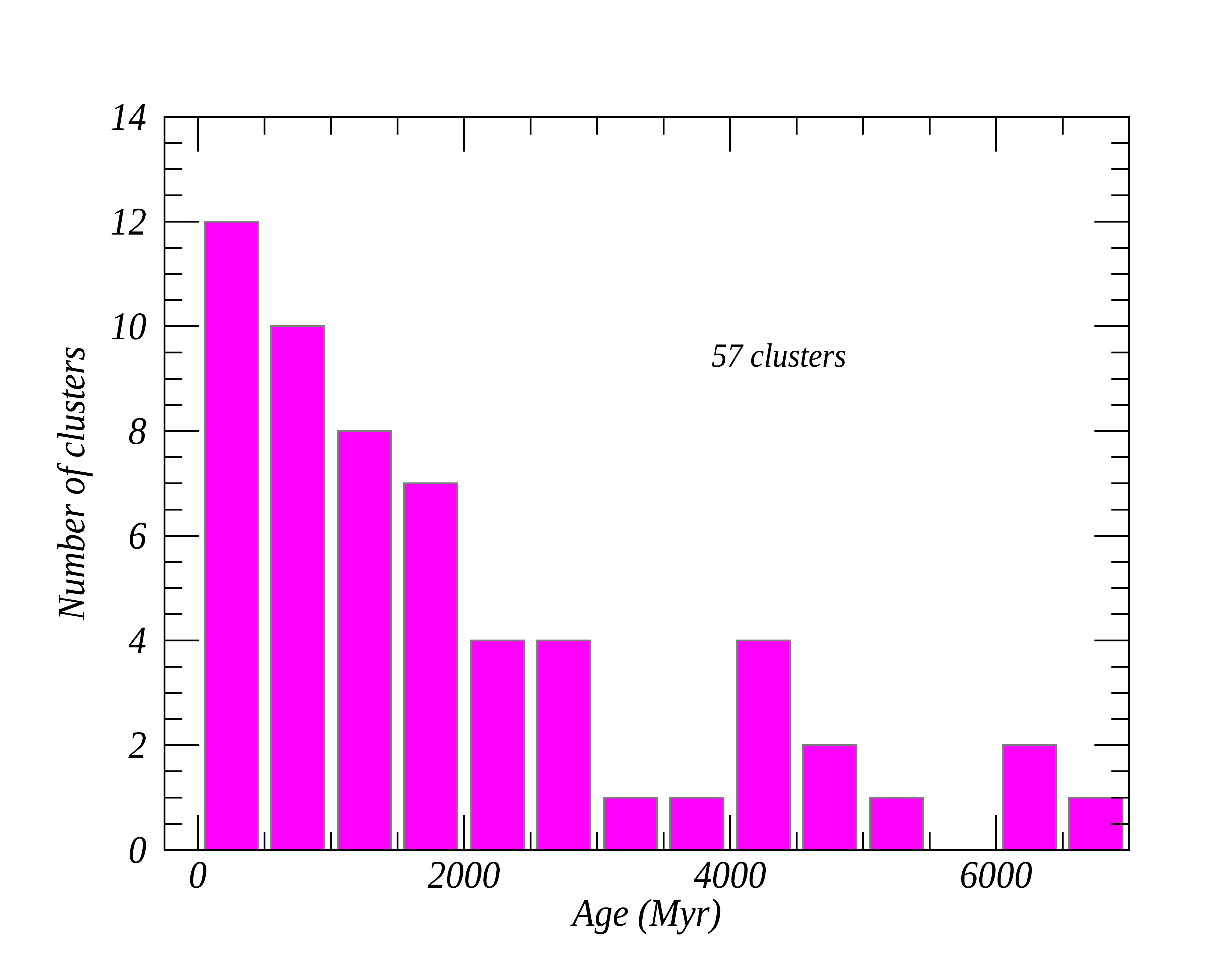}
\caption{Histogram of the ages from CG20 for our sample of open clusters (age$\,\ge\,$120~Myr). The bin size is 0.5~Gyr. } 
\label{fig:histo:cluster}
\end{figure}

The Gaia-ESO {\sc idr6} contains 86 open clusters  \citep[87 considering the two clusters in NGC2451A ans B,][]{randich18}, including also calibration ones and those retrieved from the ESO archive. Our analysis considers 57 clusters (over 62) with age $>120$~Myr  hosting evolved giant stars in which lithium abundance is available, considering our constraints on stellar parameters and Li abundances.  We excluded five clusters, in which no giant stars were observed or with a very poor membership (e.g. Loden~165).  The histogram of the age distribution of the selected clusters, determined homogeneously by \citet[][hereafter, CG20]{CG20} 
is shown in Fig.~\ref{fig:histo:cluster}. The cluster parameters are presented in Table~\ref{tab:clusters}, including: cluster name, age, distance, and galactocentric radius from CG20, mean radial velocity (RV) and [Fe/H] from the UVES members in Gaia-ESO {\sc idr6}, MSTO mass derived from the Parsec isochrones that were used by CG20 for age determination \citep{bressan12} and  the selected isochrones. 

\begin{figure}[h!]
\centering
\includegraphics[width=\hsize]{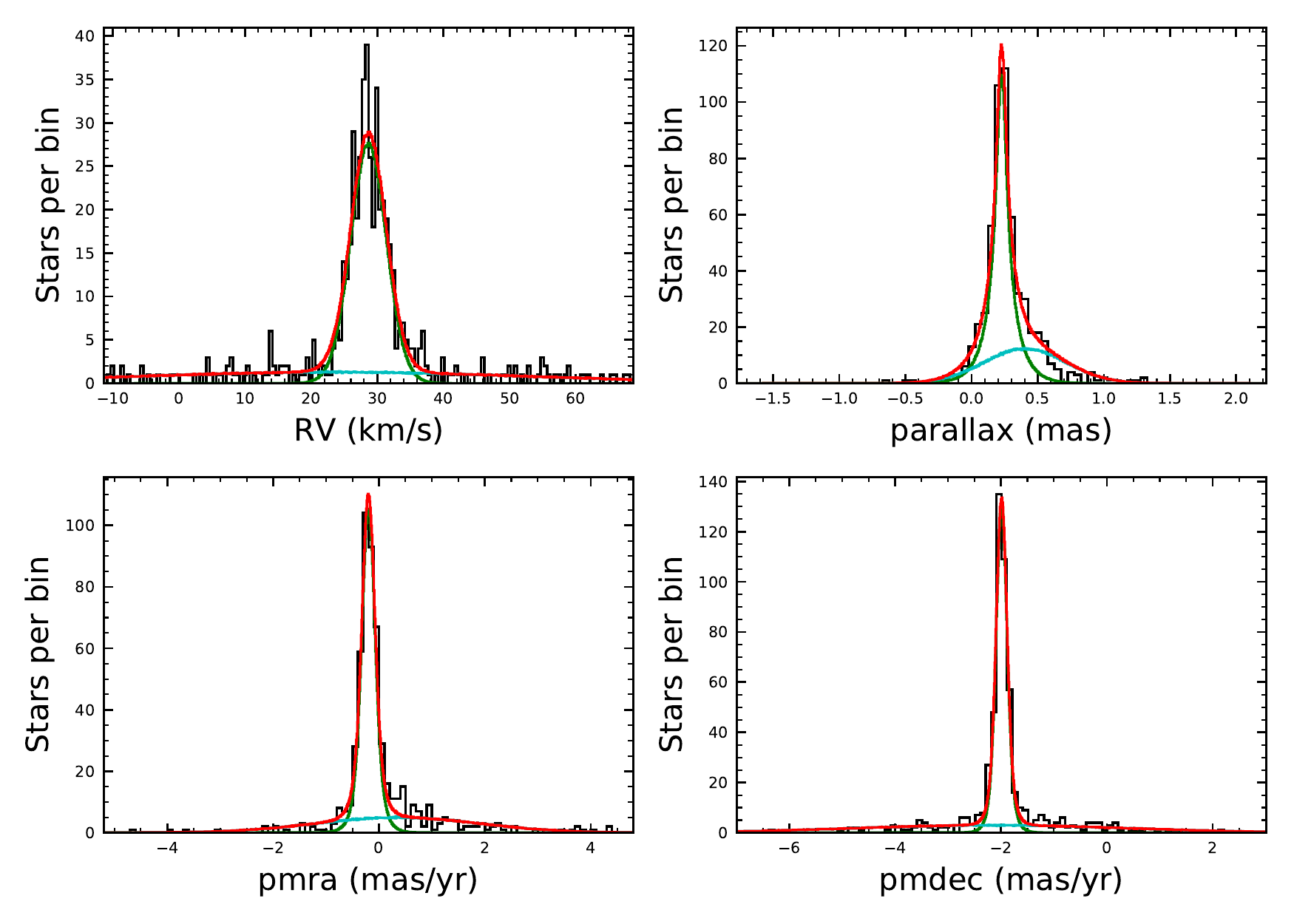}
\caption{Result of the multivariate Gaussian fit of the distribution of RV, parallax and proper motions (in RA and Dec) for NGC~2158: the total distribution is shown in red,
while the green and cyan lines show the cluster and field components, respectively.}
\label{fig:members:ngc2158}
\end{figure}

The ages of our sample clusters span from 120~Myr to about 7~Gyr. 
For the clusters containing more than 20 targets, the selection of member stars was done by performing a simultaneous fit of the Gaia-ESO RVs and of the parallaxes and proper motions from {\it Gaia} {\sc edr3} \citep{gaia20}. To this aim, we extended the method described by \citet{francio18} and \citet{roccatagliata18}, adding the RV as fourth (independent) parameter. For each cluster, the distribution was fitted with two multivariate Gaussians, one for the cluster and one for the field,
taking measurement errors and the {\it Gaia} covariance matrix into account.
When strong contamination from the field is present, we
first discarded the objects located at more than 5-$\sigma$ from a first-guess average centroid for the cluster parallax and proper motions.
{\it Gaia} astrometry was used in the fit only if  the Renormalised Unit Weight Error (RUWE), a statistical indicator of the quality of the data, is $\le 1.4$. An example of the fit is shown in Fig.~\ref{fig:members:ngc2158} for the case of NGC~2158.
We then computed a membership probability for each star in the usual way, i.e. dividing the cluster distribution by the total one, and selected as members the objects with $P>0.8$.
For the remaining clusters with less than 20 targets, to which the above method cannot be applied, we first derived the peak and standard deviation of the RV distribution, and selected stars within 2-$\sigma$ of the peak. We then computed the average parallax and proper motion and the corresponding standard deviations for the selected stars, and we further excluded those differing more than 2-$\sigma$ from the average values.
We compared the present selection with the one of CG20,
finding in general an excellent agreement.
With our selection we can add  some members among the fainter stars or in crowded fields for which CG20 do not provide a membership probability. 

\begin{figure}[h!]
\centering
\includegraphics[width=\hsize]{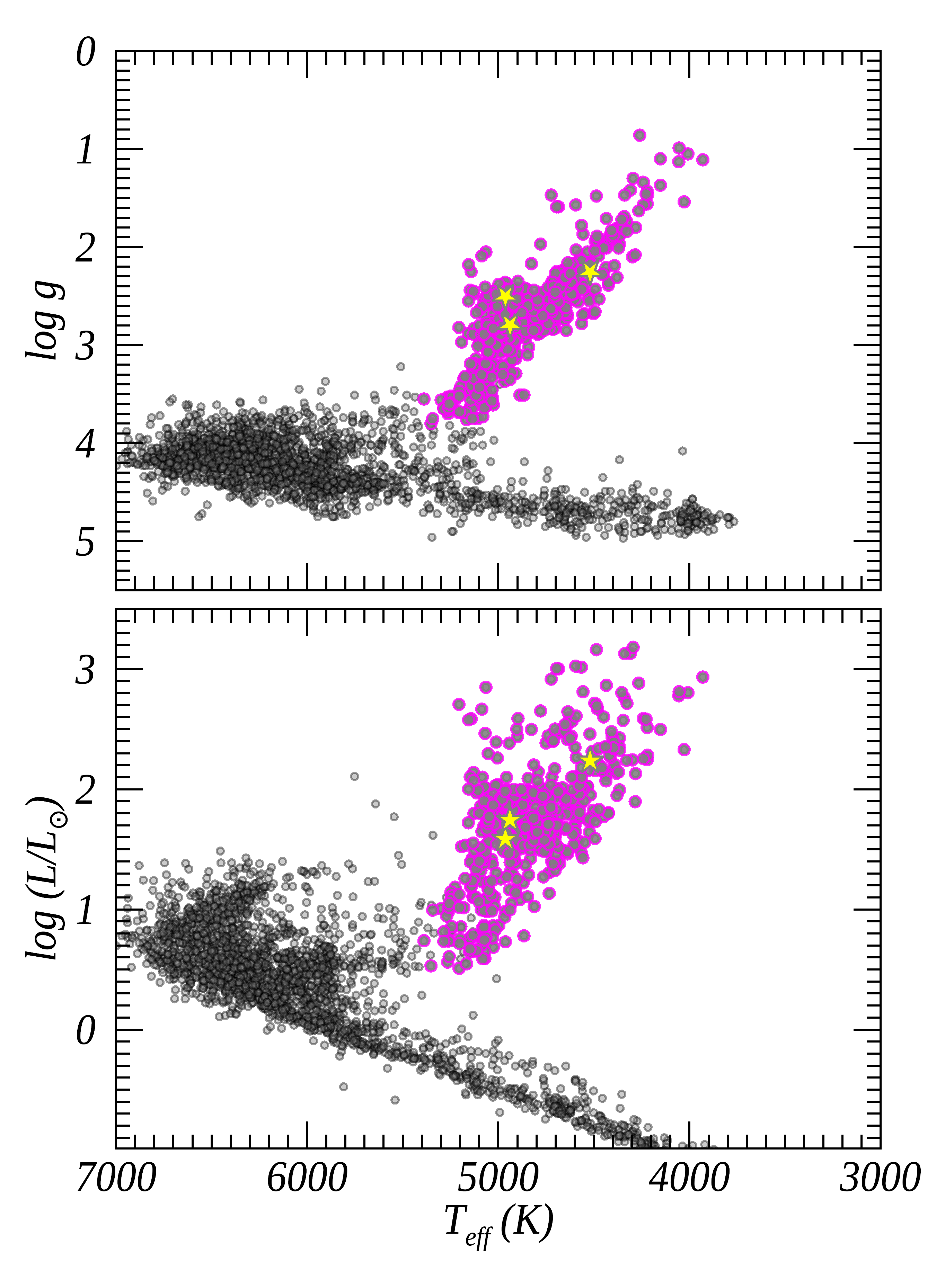}
\caption{Position in the HRD and the Kiel diagram of the members of the open clusters in Gaia-ESO {\sc idr6} with age$\,\ge\,$120~Myr selected in this study.  MS and sub-giant stars are in black, more evolved giants are in magenta. The Li-rich stars are marked with yellow stars.  
}
\label{fig:oc_giants}
\end{figure}

We made a further selection, based on the errors on the stellar parameters (error($T_\mathrm{eff}$)$<$100~K, error($\log g$)$<$0.2, error([Fe/H])$<$0.15), and including only stars with measured lithium abundances with error on A(Li) lower than 0.25~dex  or upper limits.  We relaxed the selection on the error on A(Li) for Li-rich giant stars, defined, as in \citet{casey16} and \citet{smi18}, as stars with 3800~K$\,\le T_\mathrm{eff}\le\,$5000~K, $\log g\le 3.5$ -- or $\gamma \ge$ 0.98 and log(L/L$_{\odot}$)$\,\ge\,$1~dex for star selected as giants based on their $\gamma$ index -- and A(Li)$\,\ge\,$2.0~dex,  for which we do not apply any error cut. 
The Hertzsprung-Russell diagram (hereafter HRD) and the Kiel ($\log g-T_\mathrm{eff}$) diagram of the selected members of open clusters are displayed in Fig.~\ref{fig:oc_giants}. The final number of considered cluster members is 4212 (see Table~\ref{table:summary}), of which about 18\% are giant stars.  
The stellar parameters, A(Li) (measurements and upper limits) and MSTO masses for the adopted sample of cluster stars are given in Table~\ref{table:cluster:sample}.

\subsubsection{The field star sample}
\label{sec:sample:field}

\begin{figure}[h!]
\centering
\includegraphics[width=\hsize]{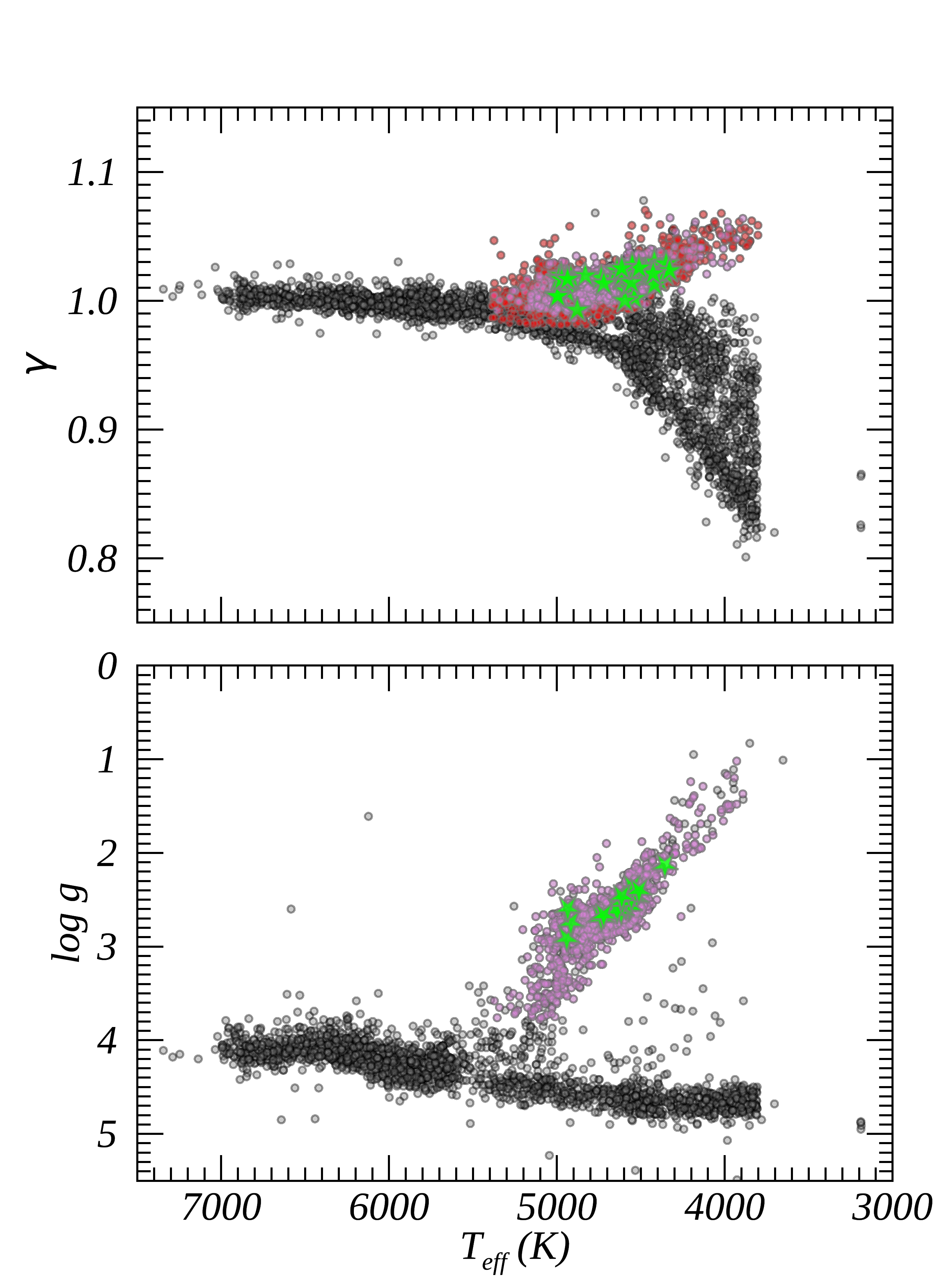}
\caption{Selection of giant stars in the Milky Way field as non members of young clusters, in the $\gamma$ versus $T_\mathrm{eff}$
diagram (upper panel) and in the Kiel diagram (bottom panel).
Giants selected on the basis of the $T_\mathrm{eff}$ and $\log g$ criterion are in pink, those selected on the basis of the
 $T_\mathrm{eff}$ and $\gamma$ criterion are in red; MS and PMS stars are marked in black.
The Li-rich giant stars are indicated with green stars. 
}
\label{fig:field_giants_young_clusters}
\end{figure}

For the sample of field stars, we select stars in two different ways, depending on the WG that analysed them. 
The first selection allows us to identify the observed field stars as non-members of young clusters with age$\,\le\,$120 Myr, which are analysed by WG12.
To select them,  we inverted the selection applied by \citet{bravi18}, keeping stars with $T_\mathrm{eff}<5400$~K and either $\gamma >0.98$, for those observed with GIRAFFE, or $\log g <3.8$  for those observed with UVES. For 4800~K$\,<T_\mathrm{eff} \le\,$5400~K we selected stars with $\gamma >0.98$, while for stars with $T_\mathrm{eff}\le\,$4800~K we adopted the selection  $\gamma\ge 1.22-5\>10^{-5}\times T_\mathrm{eff}$ to avoid contamination by the coolest MS and PMS stars.  
The selections in the $\gamma$ versus $T_\mathrm{eff}$ and $\log g$ versus $T_\mathrm{eff}$ diagrams are illustrated in the top and bottom panels of Fig.~\ref{fig:field_giants_young_clusters}, respectively.
In the figure, we also indicate  the Li-rich red giant stars with A(Li)$\,\ge\,$2.0~dex. 

\begin{figure}[h!]
\centering
\includegraphics[width=\hsize]{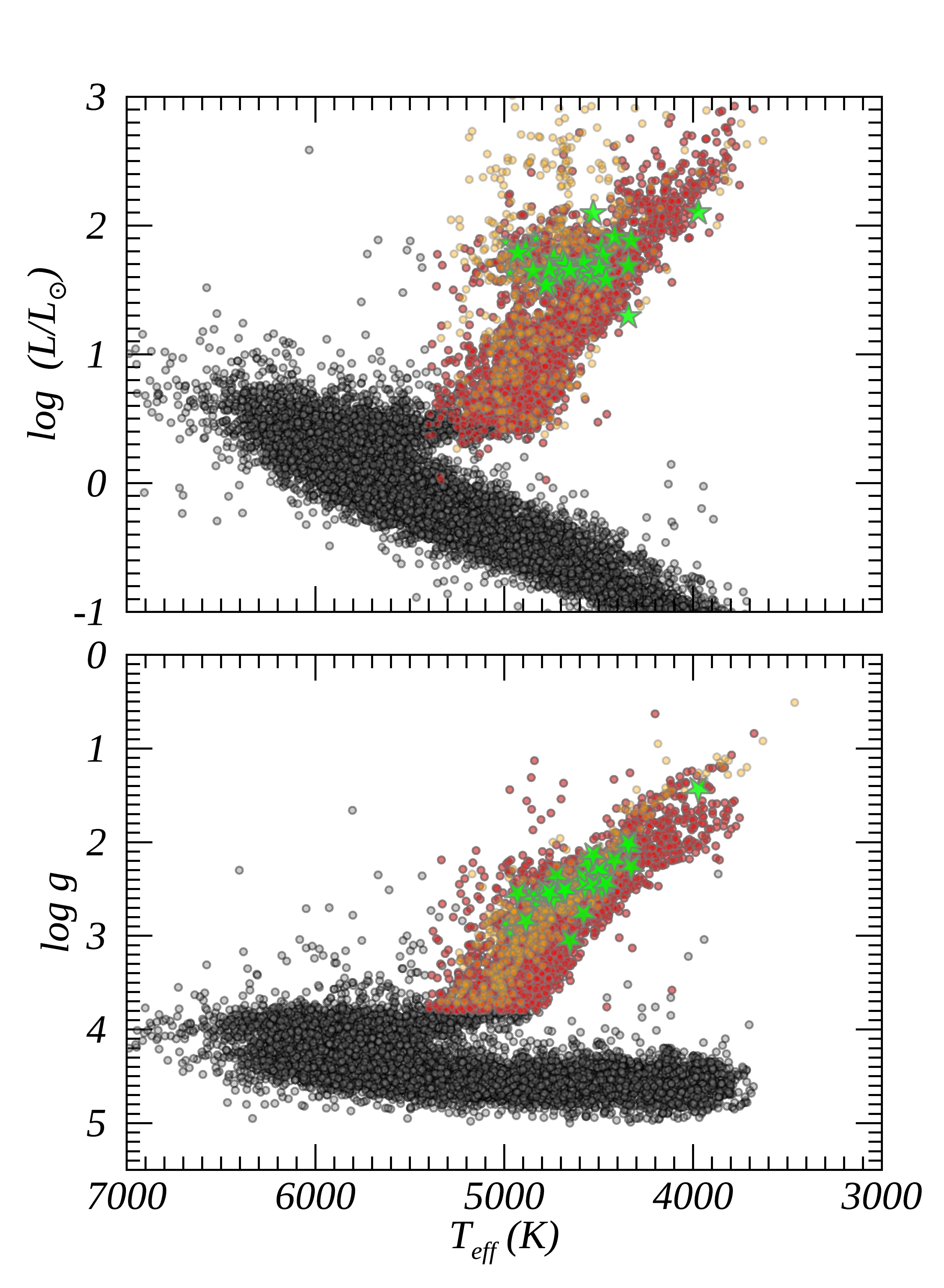}
\caption{Selection of giant stars in the Milky Way field as non member of clusters with age$>$120~Myr (orange) and in the field (red). In black,  MS and sub-giant stars.  In the upper panel the HR diagram and in the bottom panel the  Kiel  diagram. In both panels, Li-rich giant stars are indicated by green stars. }
\label{fig:field_giants}
\end{figure}

The second selection criterion allows us to pick: {\it i)} field stars which are non-members of the old and intermediate-age open clusters (age$\,>\,$120~Myr): in this selection, we take into account all stars not selected as cluster members on the basis of their radial velocities, proper motions, and parallaxes; and {\it ii)} the stars observed in the Gaia-ESO field samples, by selecting the  GES\_FLD keywords related to the field stars (GES\_MW for general Milky Way fields, GES\_MW\_BL for fields in the direction of the Galactic bulge, GES\_K2 for stars observed in  Kepler2 (K2) fields, GES\_CR for stars observed in CoRoT fields). 
We combined the two samples of field stars, performing a further selection on stellar parameter uncertainties as done for the sample of stars in open clusters. 
The results of the selection are shown in Fig.~\ref{fig:field_giants}, where Li-rich giant stars are also indicated.  

\begin{table}
\caption{Summary of the selected samples.}
\label{table:summary}
\centering
\begin{tabular}{lll}
\hline\hline
\multicolumn{1}{c}{Sample} &
\multicolumn{1}{c}{Stars with A(Li)} &
\multicolumn{1}{c}{Detections} \\
\hline
Gaia-ESO {\sc idr6}                         & 38090 & 27256 \\
Gaia-ESO {\sc idr6} + {\it Gaia} {\sc dr3}  & 37940 & 27142 \\
Field stars                     & 7369  & 3866   \\
Cluster members                 & 4212 & 3497 \\
\hline\end{tabular}
\end{table}

To improve the quality of the sample, we cross-matched our catalogue with {\it Gaia} {\sc edr3} and selected only stars for which  the parallaxes have uncertainties within 10\%.
For them,  we computed the stellar luminosity using the {\it geometric} distances from \citet{bailerjones20} and the 
{\it Gaia} {\sc edr3} $G$ magnitudes, converted into V magnitudes using the $G_\mathrm{BP}$ and $G_\mathrm{RP}$ colours. 
We computed the bolometric magnitudes using the bolometric corrections BC(K), based on the V$-$K colours\footnote{K magnitudes are obtained from the 2MASS catalog \citep{2mass}}, from the tables provided by \citet{alonso99} for dwarf and giant stars. 
We used bolometric corrections based on colour instead of more recent calibrations, as the one of \citet{casagrande18}, based on stellar parameters, in order to provide corrections independent of stellar parameters, which can have considerably large uncertainties, especially for the GIRAFFE spectra, since the spectral range of HR15N is not optimised to derive precise atmospheric parameters, and to be able to apply them also to stars for which we have $\gamma$ instead of $\log g$. 
We adopted the reddening values from the 3D extinction map of \citet{green19}, extracting E(B-V) in the line-of-sight and at the distance of each star, when available, and from the 2D extinction map of \citet{schlegen98} in the remaining cases.   
As expected, E(B-V) from \citet{schlegen98} is typically larger than E(B-V) from \citet{green19}, the latter being integrated over larger distances.  We take as an estimate of E(B-V) the minimum of the two values, which is equivalent to using \citet{green19} when it is available, and exclude stars with E(B-V)$>$1. For Li-rich giant stars, we relaxed the selection on parallax, removing the cuts on the parallax relative error, to avoid losing some of them. 
The final sample for which we have high-quality luminosities from {\it Gaia}, with error on $\log (L/L_{\sun})$ lower than 0.15~dex, contains 7368 stars, of which about 56\% are giant stars (see Table~\ref{table:summary}).   
The stellar parameters (including $\gamma$), A(Li) (measurements and upper limits) and masses for the adopted sample of field stars are given in Table~\ref{table:field:sample}.

\begin{figure}[h!]
\centering
\includegraphics[width=\hsize]{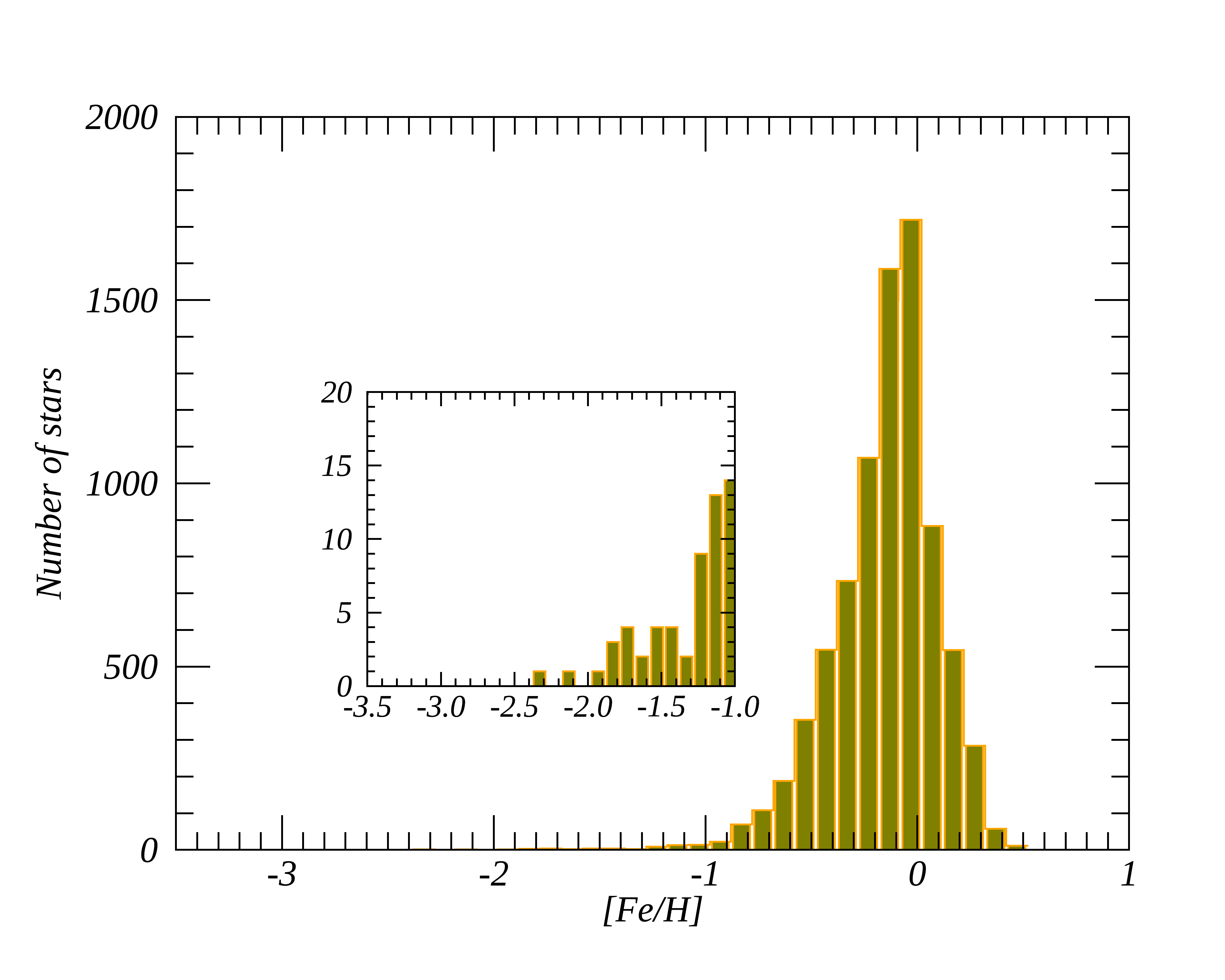}
\caption{Histogram of [Fe/H] for the sample of field stars.  In the insert, we show the tail at low [Fe/H]. 
}
\label{fig:histo}
\end{figure}

The histogram of the distribution of [Fe/H] for the field stars is shown in Fig~\ref{fig:histo}.
The peak of the metallicity distribution function (MDF) is at [Fe/H]$\,\sim -0.1$, with a tail of lower metallicity stars down to [Fe/H]$\,\sim -1.0$,
and of higher metallicity stars reaching [Fe/H]$\,\sim +0.5$. 
In the insert of Fig~\ref{fig:histo} we highlight the low metallicity tail of the MDF: there are some stars with $-2.0<\,$[Fe/H]$\,<-1.0$ and a few ones below [Fe/H]$\,=-2.0$. 

 \section{Comparison with other catalogues}
 \label{section:cataloguecomparison}

\begin{figure}[h!]
\centering
\includegraphics[width=\hsize]{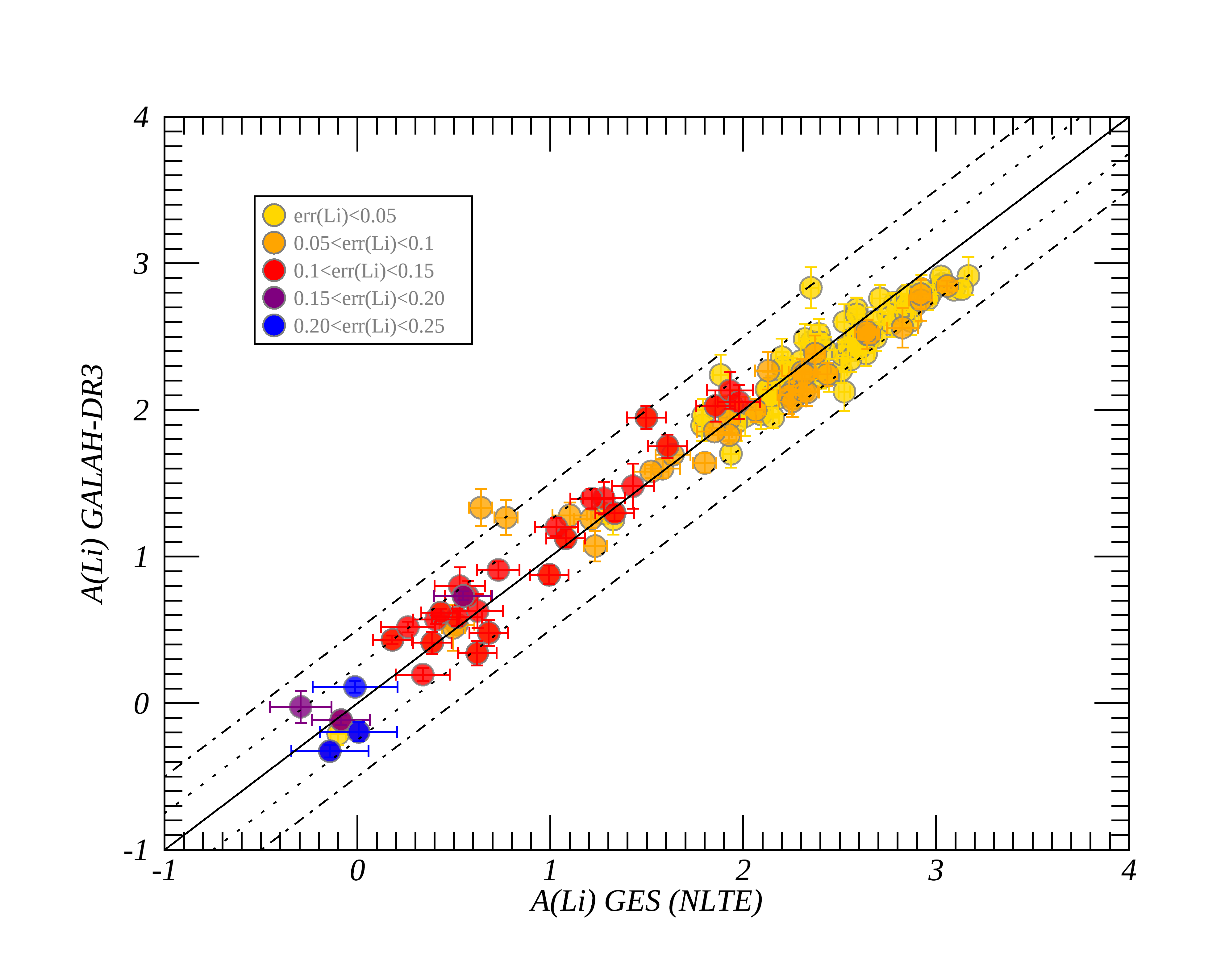}
\caption{A(Li) corrected for 3D-NLTE effects for stars in common between Gaia-ESO {\sc idr6} (not all of them being part of our final selection) 
and GALAH {\sc dr3} \citep{buder20}. The circles are colour-coded by the error from Gaia-ESO in A(Li).
The continuous-line is the 1-to-1 relation, while the two dashed lines are  at $\pm$0.5~dex and the two dotted lines at $\pm$0.25~dex. 
}
\label{fig:galah}
\end{figure}


Gaia-ESO {\sc idr6} has 676 stars in common with GALAH {\sc dr3} \citep{buder20} with available Li abundances in both surveys.
These stars belong to several different groups (not all being part of the selection discussed in the present paper): the open clusters $\lambda$ Ori, 25 Ori and Cha~I,  NGC~2243,  NGC~2516, IC~4665, M~67, NGC~6253, Rup~147, Rup~7, Trumpler~20,  the globular clusters NGC~362 and NGC~104, 
several Milky Way field stars (mainly turn-off stars), and stars in the CoRoT and K2 fields. 
Among the sample of  stars in common, we select a sub-sample of 135 stars, with high quality A(Li) in both surveys: for Gaia-ESO we take only Li measurements with error lower than 0.25~dex and for which we can compute NLTE corrections following \citet{wang21} for a meaningful comparison with GALAH NLTE Li abundances; from GALAH we consider Li abundances with flag\_sp=0 and flag\_li\_fe=0  (quality flags that indicate the good quality of the spectral analysis and of the A(Li) determination, respectively), and error lower than 0.25~dex. 
We recall that not all of the 135 stars are used in the present work, but they are included here for a general comparison between the two surveys.   
The comparison is shown in Fig.~\ref{fig:galah}, 
with the abundances colour-coded by the Gaia-ESO uncertainties. 
The lithium abundances in the two surveys agree very well. 
There is some scatter in the comparison, but in most cases the agreement between Gaia-ESO and GALAH is within 0.25~dex.

We also compare stars in common between Gaia-ESO {\sc idr6}, to which we apply the NLTE corrections from \citet{wang21}, and the AMBRE sample of \citet{Guiglion16}, considering the Li abundances corrected  by NLTE effects  \citep{lind09} available in \citet{Guiglion16}. The agreement is quite good, with a small offset towards higher A(Li) in the \citet{Guiglion16} sample. 

\section{Post-MS lithium evolution: comparison with stellar model predictions}
 \label{section:datamodelcomparison}

The results of the recent studies using {\it Gaia} and  large spectroscopic surveys \citep[e.g.][]{deeppak19, deepak20, c20, kumar20a, yan21} have been instructive for the understanding of the Li post-MS evolution. 
With the present work, we provide additional information for a better understanding of  the  macroscopic magnetic hydrodynamic (MHD) transport processes acting along the evolution of low- and intermediate-mass stars at different metallicities.
With the Gaia-ESO results, we can indeed expand the analysis to a larger sample of stars in different metallicity ranges, thus investigating the Li evolution
from the MSTO to the RGB sequence in different conditions. In particular, the combination of field stars and of members of open clusters allow us to cover a larger range of stellar masses, from the lowest masses in the field and in the old clusters, to the highest masses in the young open clusters.

In the following sections, we compare our results with a set of stellar models from \citet{lagarde12}. 
The models are computed with the stellar-evolution code {\sc STAREVOL} (v3.00). 
The mechanisms  included for the transport of chemicals are: {\it i)} the standard mechanism due to convection; {\it ii)} the thermohaline double-diffusive instability (so-called thermohaline mixing) that is expected to develop in low-mass stars along the RGB at the luminosity bump and in intermediate-mass stars on the early-AGB; it takes place when, in the external part of the hydrogen-burning shell around the degenerate stellar core, there is an inversion in the mean molecular weight gradient in a thermalised medium   \citep{cczahn07, CL10}; {\it iii)} rotation-induced mixing ``\,\`a la Zahn'' with the vertical and horizontal turbulent coefficients from \citet{TZ1997} and \citet{JPZ1992} respectively, computed considering, at the zero-age MS (ZAMS), a rotational velocity equal to 30\% of the critical one, which means typical velocities on the MS between 90 and 137~km~s$^{-1}$. 
In our comparison, we consider both the classical models, in which only mixing due to convection is applied, and the models in which the effect of rotation-induced mixing and thermohaline mixing are included. 
We recall that these kinds of models are also crucial to explain the behaviour of Li abundance during MS. Among the various considered additional mechanisms, both rotation \citep[see][]{SR05} and overshooting mixing \citep{CD11, zhang12} have been introduced to reproduce the observational properties of the clusters and at the same time the properties of the Sun. 
However, for low mass solar-type stars with relatively extended convective envelopes,  hydrodynamic processes induced
by rotation, as, for instance,  meridional circulation and shear mixing, predict large rotation gradients within the interior, needing, e.g., internal gravity waves or other  mechanisms, as penetrative convection, tachocline mixing, and additional turbulence, to explain both the rotation profile and the surface abundance of lithium in solar-type stars of various ages \citep[see][]{CT05, 2021A&A...646A..48D}.

\subsection{Li evolution in field stars}
\label{subsection:Lifield}

\begin{figure*}[h!]
\centering
\includegraphics[width=\hsize]{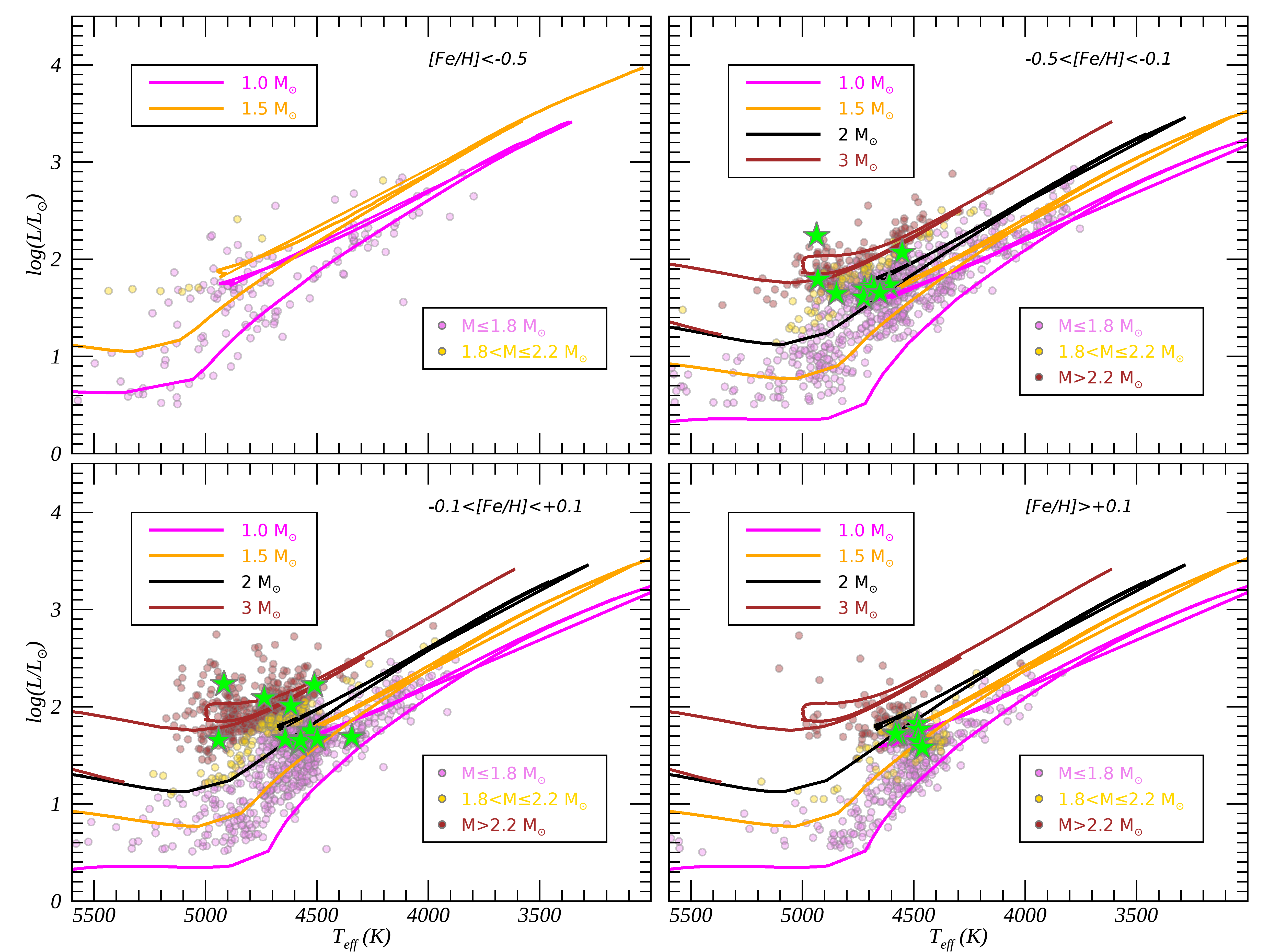}
\caption{Location of our  sample of sub-giant and giant field stars in the HR diagram in the four metallicity bins. The  stars (filled circles) are colour-coded by their masses: in pink stars with $M\le 1.2 \,M_\sun$, in yellow stars with $1.2 \,M_{\sun} < M \le 2.2 \,M_{\sun}$, and in red stars with $M > 2.2 \,M_{\sun}$.
The Li-rich giant stars with a mass determination are marked with green stars. The theoretical evolutionary tracks are plotted for masses between 1 and 3~$M_{\sun}$. In the  top-left panel, corresponding to the metallicity bin with [Fe/H]$\,\le -0.5$, we adopt the tracks computed for [Fe/H]$\,=-0.56$, while in the other bins we plot the tracks at the solar metallicity. 
} 
\label{fig:HR:tracks}
\end{figure*}

Since stellar masses play a fundamental role also during the post-main sequence evolution, we estimate them to compare with the appropriate theoretical models from \citet{lagarde12}. 
Masses of field stars are computed using a maximum-likelihood technique described in \citet{c20}, and adapted from \citet{valle14}, comparing $T_\mathrm{eff}$, luminosity, and [Fe/H] of individual stars to the theoretical evolutionary tracks of \citet{lagarde12}. The errors on the three parameters are taken into account to estimate the uncertainty on the stellar mass. 
In Fig.~\ref{fig:HR:tracks} we show the HR diagram of our sample of giant and sub-giant field stars (see Section~\ref{sec:sample} for a definition), divided in four metallicity bins, and overlaid on the corresponding evolutionary tracks from \citet{lagarde12}. Most of our stars are located in the ascending and upper parts of the RGB and in the clump. 

\begin{figure*}[h!]
\centering
\includegraphics[width=\hsize]{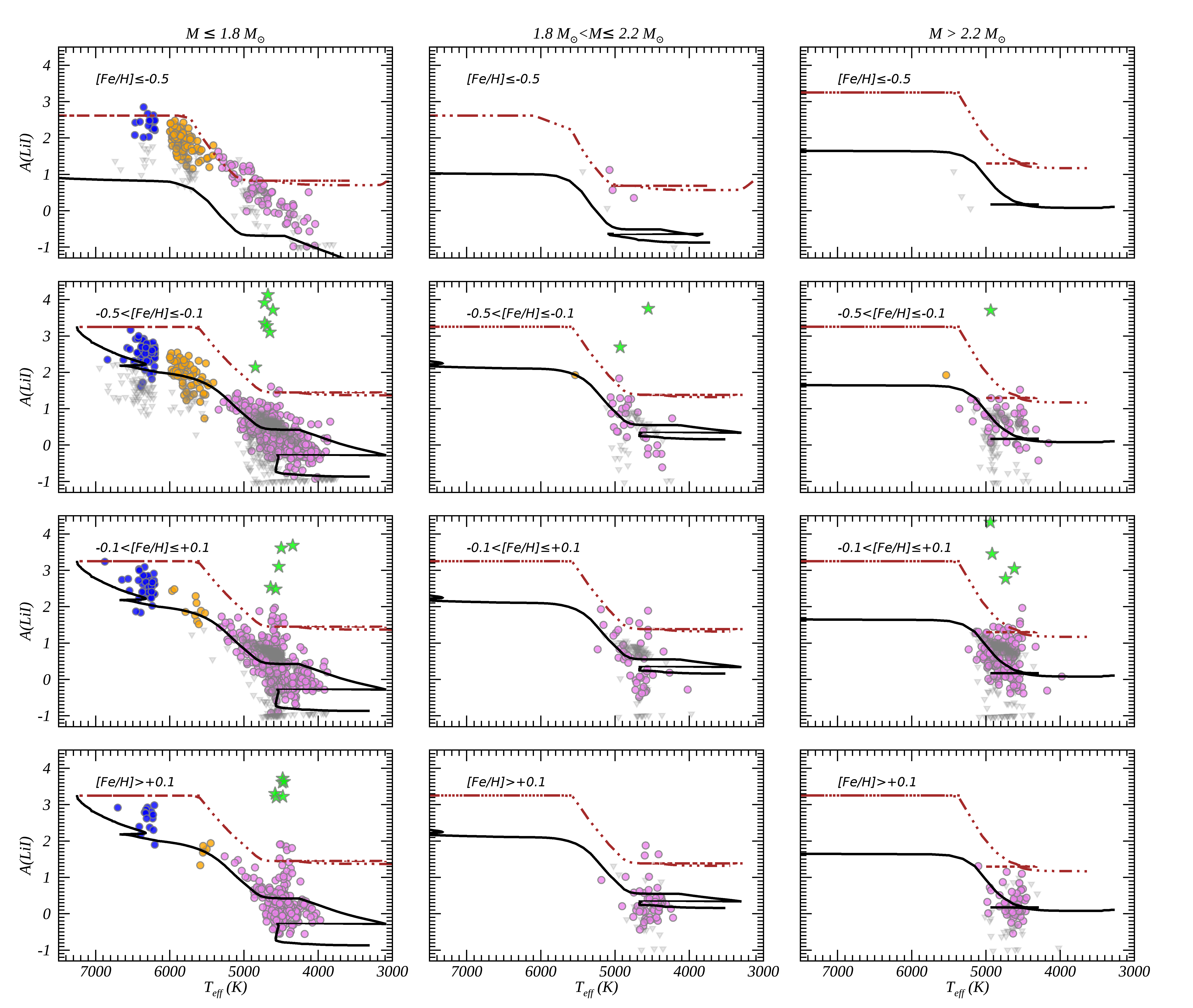}
\caption{A(Li) vs. $T_\mathrm{eff}$ in the field stars sample for which the mass was estimated. We plot in blue the MSTO stars (with $T_\mathrm{eff}> 6200$~K), in orange the sub-giant stars,  and in pink the giant stars. Li upper limits are shown with grey triangles. The curves are the predictions of the models of \citet{lagarde12} for 1.5~$M_{\odot}$, 2.0~$M_{\odot}$, and 3.0~$M_{\odot}$ with standard mixing (dashed lines) and with rotation-induced mixing and thermohaline instability (continuous lines). In the panels for [Fe/H]$\,<-0.5$~dex and M$\leq$2.2~$M_{\odot}$, we adopt the models at [Fe/H]$\,=-0.56$, in the other panels the models at [Fe/H]$\,=0$.
The Li-rich stars with a mass determination are indicated with green stars.}
\label{fig:li1}
\end{figure*}

In Fig.~\ref{fig:li1}, we show the evolution of A(Li) as a function of the  effective temperature in three mass bins: 
$M\le 1.8\,M_{\odot}$, $1.8\,M_{\odot}<M\le 2.2\,M_{\odot}$, and $M>2.2\,M_{\odot}$.
Each mass bin is further divided in the four  metallicity bins:
[Fe/H]$\,\le -0.5$, $-0.5<\,$[Fe/H]$\,\le -0.1$, $-0.1<\,$[Fe/H]$\,\le +0.1$, and  [Fe/H]$\,>+0.1$. 
In the panels, we show both sub-giant and giant stars. When available in our samples, we include also MSTO stars (at the corresponding mass and metallicity) with $T_\mathrm{eff}>6200$~K which might have preserved their initial Li as expected in the classical models.  
We compare the observations with the theoretical predictions of the models of \citet{lagarde12} (classical  and with thermohaline and rotation-induced mixings).

We see in Fig.~\ref{fig:li1}  that the Li surface evolution predicted by classical  models is very similar for stars of different masses and metallicities. In that case, the surface Li depletion
is only due to the FDU, which starts at  $T_\mathrm{eff}$ around 5600~K. Since for low-mass stars the maximum depth reached by the base of the convective envelope during the FDU is almost independent of the stellar mass,  A(Li) reaches similar values for all models of same metallicity: A(Li)$\sim\,1.3-1.6$~dex  at solar metallicity, and A(Li)$\sim\,0.7-1.0$~dex for the sub-solar models at [Fe/H]$\,=-0.56$. 
These values are in agreement with previous theoretical studies \citep[starting e.g. with the early work of][]{1967ApJ...147..624I}. 
After the end of the FDU (around $T_\mathrm{eff}\sim$ 5000 to 4500~K, depending on the mass and metallicity), the convective envelope withdraws in mass, and no more surface Li depletion is expected. 

As already mentioned in the introduction (see references to previous studies in Sect.~\ref{section:introduction}), and as evidenced in Fig.~\ref{fig:li1}, 
these classical predictions do not reproduce the observed Li behaviour. 
Indeed, 
for the masses and metallicities explored here, 
Li depletion starts earlier (i.e., at higher $T_\mathrm{eff}$) on the subgiant branch than predicted by the classical models, and is more efficient. Additionally, the second depletion episode that occurs in stars with masses below $\sim\,2.2\,M_{\odot}$ after the so-called RGB luminosity bump (at $T_\mathrm{eff}\sim\,$4200 K, i.e., when the H-burning shell has passed the chemical discontinuity left behind by the FDU) is not predicted either by the classical stellar evolution theory.

The Li data for field stars presented in Fig.~\ref{fig:li1} thus confirm the need of including in stellar evolution models rotation-induced mixing over the entire mass and metallicity range considered, as well as thermohaline mixing in low-mass stars that pass through the RGB bump before igniting He in their degenerate core at the tip of the RGB. On one hand, rotation-induced mixing changes the  abundances profile in the stellar interiors already during the main sequence, enlarging the size of the Li-free region. 
As a consequence, compared to classical predictions, surface Li depletion starts earlier (i.e., at higher $T_\mathrm{eff}$), 
and lower Li abundances are predicted after the end of the FDU \citep{palacios03,c20}.

Figure~\ref{fig:li1} clearly shows that rotating models reproduce the Li abundance from the MSTO on, whatever the mass and metallicity ranges.  
On the other hand, the introduction in the models of the thermohaline double diffusive instability as proposed by \citet{cczahn07} reconciles the theoretical predictions with the Li data in the brightest and coolest RGB stars. Indeed, when these low-mass evolved stars pass the RGB bump, this instability develops between the base of the convective envelope and the Li-burning regions, because of the mean molecular weight inversion resulting from the $^3$He($^3$He, 2p)$^4$He reaction in the Hydrogen burning shell. 

As already shown with other samples from the literature \citep{cczahn07,CL10,lagarde15,c20}, this explains the second drop of the surface Li abundance highlighted by the Gaia-ESO field star data shown in Fig.~\ref{fig:li1}.
Inside stars more massive than $\sim\,2.2\,M_{\odot}$, however, the thermohaline instability does not set in because they do not pass through the RGB bump, as they ignite central helium-burning earlier in non-degenerate conditions. For this mass range (panels in the right column of Fig.~\ref{fig:li1}) the lowest Li abundances observed are well explained by rotation alone, as discussed before.  

Last but not least, we see in Fig.~\ref{fig:li1} that there is a conspicuous number of giant stars with an anomalously high A(Li) with respect to the general trend discussed above.
Some of them, from the previous Gaia-ESO data releases, have been studied  in detail by \citet{casey16} and \citet{smi18}. We discuss them in Sect.~\ref{sec:lithium:rich}.

\subsection{Li evolution in open clusters}
\label{subsection:Liopenclusters}

Lithium abundances in open clusters provide an effective way to probe mixing processes in stars of different masses and metallicity, following them along the different evolutionary sequences. 
Since the ages of open clusters can be derived with good accuracy from the isochrone fitting  
of their whole evolutionary sequence, we can estimate the masses of their evolved stars by assuming them to be those of the MSTO stars. Moreover, high-resolution spectra provide a detailed chemical composition for the cluster, which has usually a high level of homogeneity \citep[see, e.g.][]{desilva06, carrera13, bovy16, liu16}.  

In this way, stars in open clusters might effectively serve to study the changes in Li abundance during  post-MS evolution, in samples with similar masses and metallicity.
Several works have been dedicated to the study of the evolution of Li abundance along specific parts of the colour-magnitude diagrams (CMD) of open clusters. Some works \citep{randich02, randich07, smi10, canto11, pace12, anthony18, Deliyannis19} studied Li abundance in MS, sub-giant stars, and RGB stars in several open clusters, finding that non-standard mixing processes are needed to explain the observed trends. Other papers focused on Li in RGB stars and on the occurrence of Li-rich giants in open clusters \citep[see e.g.][]{at12,monaco14, delgado16, krolikowski16, aguilera16, carlberg15, carlber16}. 

In this work, we present a large sample of stars, members of 57 open clusters,  with 0.12~Gyr$\,<\,$ages$\,<$\,7~Gyr, spanning from the inner disc to the outer Galaxy, with R$_{\rm GC}$ in the range $\sim\,$6--20~kpc, and with metallicities $-$0.44~dex$\,<\,$[Fe/H]$\,<+0.27$~dex. 

\begin{figure*}[h!]
\centering
\includegraphics[width=\hsize]{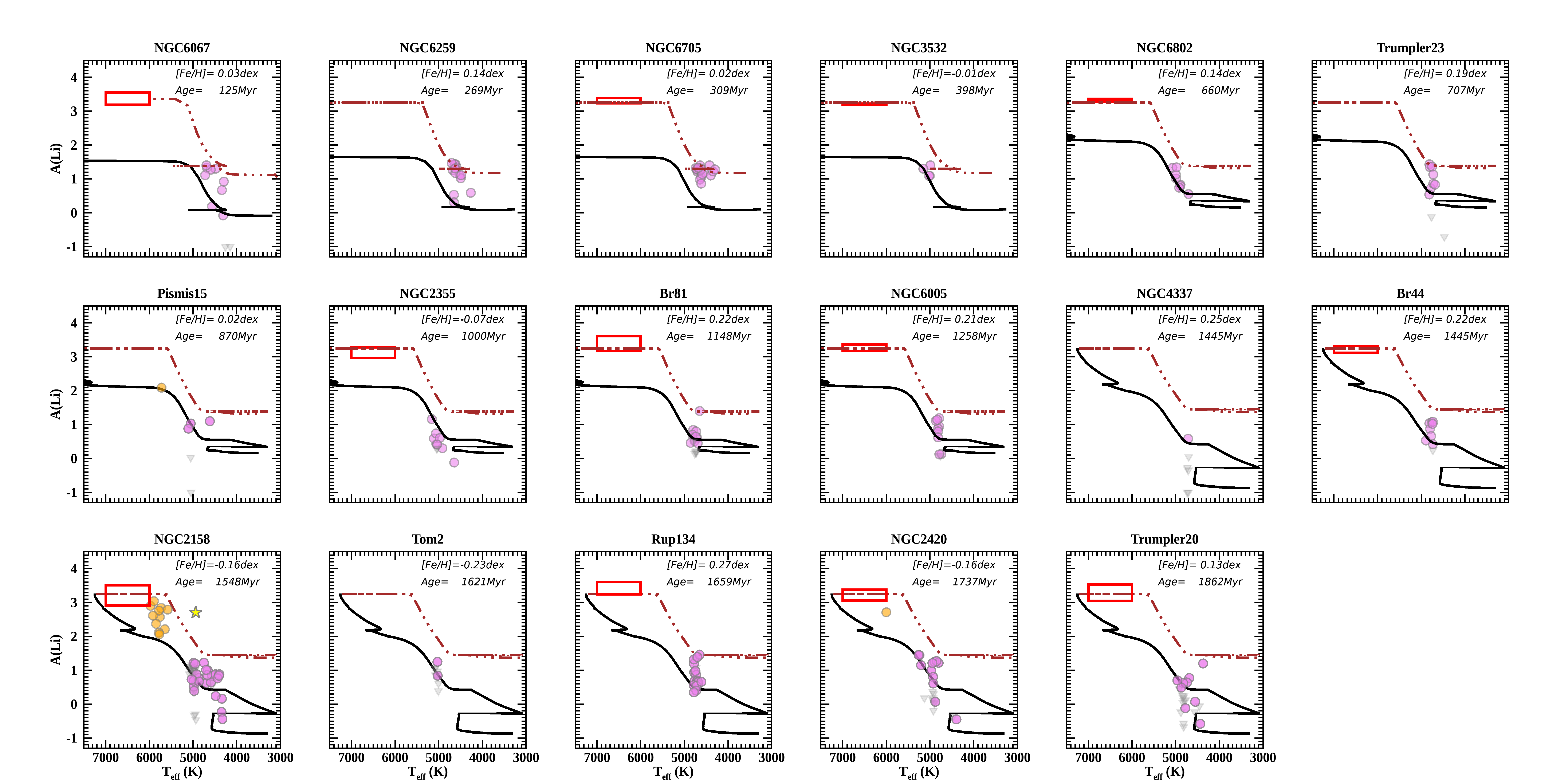}
\caption{A(Li) versus $T_\mathrm{eff}$ for 17 Gaia-ESO clusters with ages $\le\,$1.8~Gyr. Giant stars are indicated with pink circles, while stars with upper limits of A(Li) with grey triangles, sub-giants are indicated with orange circles, Li-rich giants are marked with yellow stars. The red rectangles show the location of the initial A(Li) derived as in \citep[][see text for details]{randich20}.  
The theoretical tracks of \citet{lagarde12} are selected on the basis of the age and metallicity of each cluster (classical models in dot-dashed brown curves, and rotation-induced mixing models with continuous black curves). Cluster metallicity and age are reported in each sub-panel.}
\label{fig:cluster1}
\end{figure*}

\begin{figure*}[h!]
\centering
\includegraphics[width=\hsize]{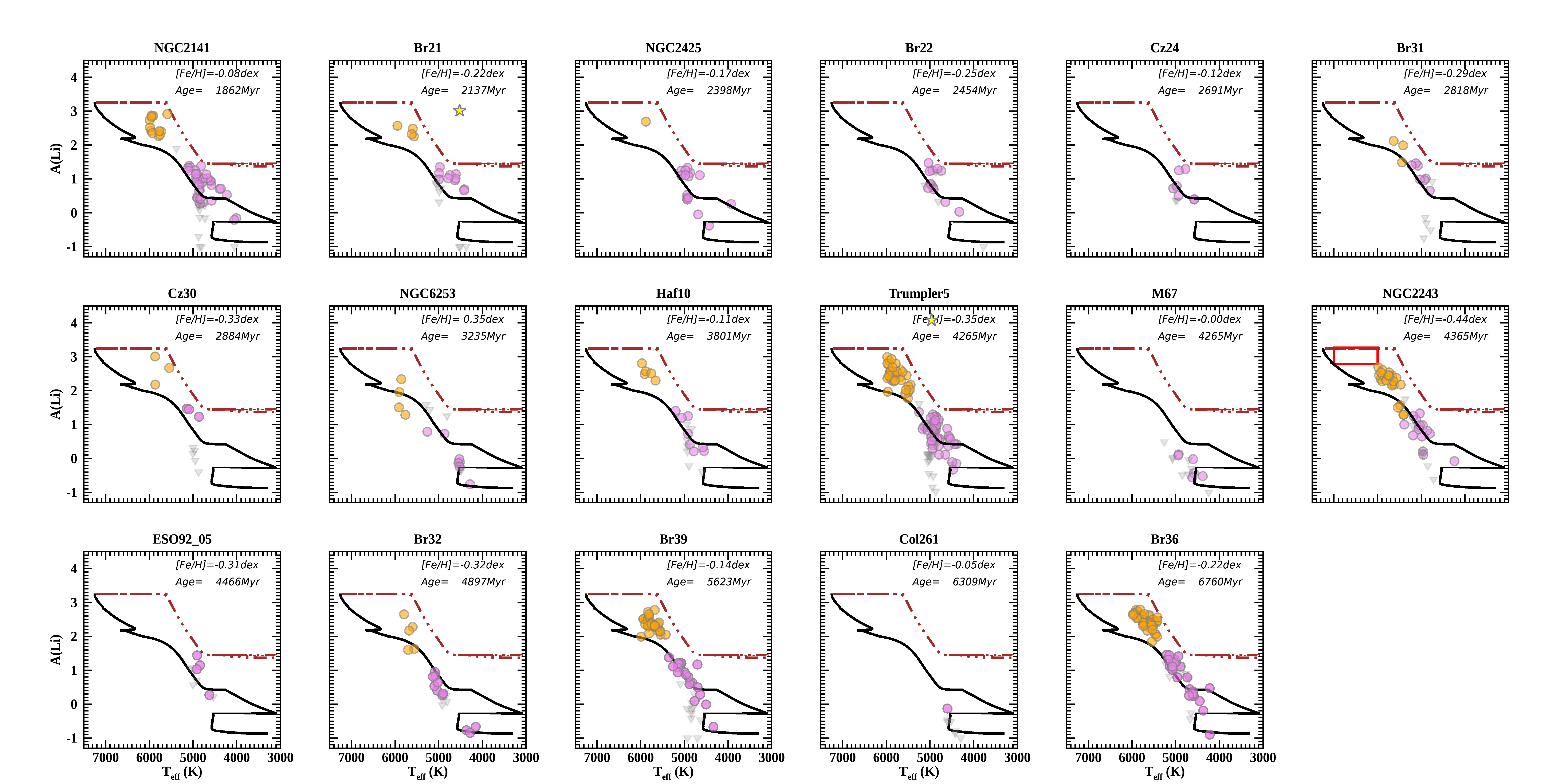}
\caption{A(Li) versus $T_\mathrm{eff}$  for the remaining Gaia-ESO clusters with age $\ge\,$1.8~Gyr. Colours and symbols are as in Fig.~\ref{fig:cluster1}. }
\label{fig:cluster2}
\end{figure*}

In Figs.~\ref{fig:cluster1} and \ref{fig:cluster2} we plot A(Li) versus $T_\mathrm{eff}$ for the 34 clusters of our sample in which A(Li) was measured in, at least, six giant stars. 
When possible, for clusters with ages between 120 and 2000~Myr we indicate the initial A(Li) that we derive from the analysis of upper MS stars located on the blue side of the so-called Li dip, following the methodology described in \citet{randich20}. For NGC~2420 and NGC~2243, given their ages, stars on the blue side of the dip are located at the upper TO \citep[see also][]{francois13}. Since they may have started to experience some post-MS Li dilution, the measured  Li is possibly a lower limit to their initial value.  
We compare the observations with the evolutionary tracks from the models of \citet{lagarde12}, as for the field stars. For each cluster we select the most appropriate model in terms of stellar mass, using the MSTO masses in Table~\ref{tab:clusters}. 
We adopt models at solar metallicity for all clusters, since the cluster [Fe/H] are closer to the solar ones than to the next metallicity  in the grid of \citet{lagarde12}.  

In Fig.~\ref{fig:cluster1}, for the youngest cluster NGC~6067 (age=130 Myr), the RGB stars reach A(Li)$\sim$1~dex, in between  the tracks of the classical model and the model with rotation at 4~$M_{\odot}$ (these models have been scaled of $-0.6$~dex in A(Li) to match the initial Li of the cluster). 
In slightly older clusters with  190~Myr$\,<\,$age$\,\le\,$400~Myr, namely NGC~6259, NGC~6705 and NGC~3532, the lithium in RGB stars reaches A(Li)$\sim$1--1.2~dex, following the track of 3~$M_{\odot}$, in between the classical ones and those with rotation. 
For the clusters in Fig.~\ref{fig:cluster1} with age in the range 400~Myr$\,<\,$age$\,\le\,$1400~Myr, from NGC~6802 to NGC~6005, in the RGB stars A(Li) settles between 0~dex and 1~dex.
These clusters are compared with tracks of 2~$M_{\odot}$ stars, where the effects of  both rotation-induced mixing and thermohaline instability are required to reproduce the decline of A(Li) with decreasing $T_\mathrm{eff}$.  
Clusters ages in the interval 1400~Myr$\,<\,$age$\,\le\,$6800~Myr are compared with tracks for 1.5~$M_{\odot}$ stars (starting from NGC~4437 in Fig.~\ref{fig:cluster1} to Berkeley~36 in Fig.~\ref{fig:cluster2}).
For all these clusters, the data are also better reproduced by the models including the effects of rotation-induced mixing and thermohaline instability.
Starting from clusters with age$\,>\,$1500~Myr (NGC~2158),  A(Li) in RGB stars reaches lower values, down to A(Li)$\,\sim\,-1$~dex, due to the increasing efficiency of the thermohaline instability. 
However for the lowest-mass range, corresponding to ages $>\,$4000~Myr, the comparison should be taken with caution. Indeed in the \citet{lagarde12} models the transport of angular momentum is driven by meridional circulation and turbulence only, while an additional transport is required to explain the internal rotation profile of low-mass stars, both on the MS \citep[see references in][for the case of the Sun and solar-type stars]{2021A&A...646A..48D} and along the red giant branch \citep[e.g.][and references therein]{2019A&A...621A..66E}. For this reason, for the oldest clusters we compare the observations with the model for 1.5~$M_{\odot}$.

\begin{figure*}[h!]
\centering
\includegraphics[width=\hsize]{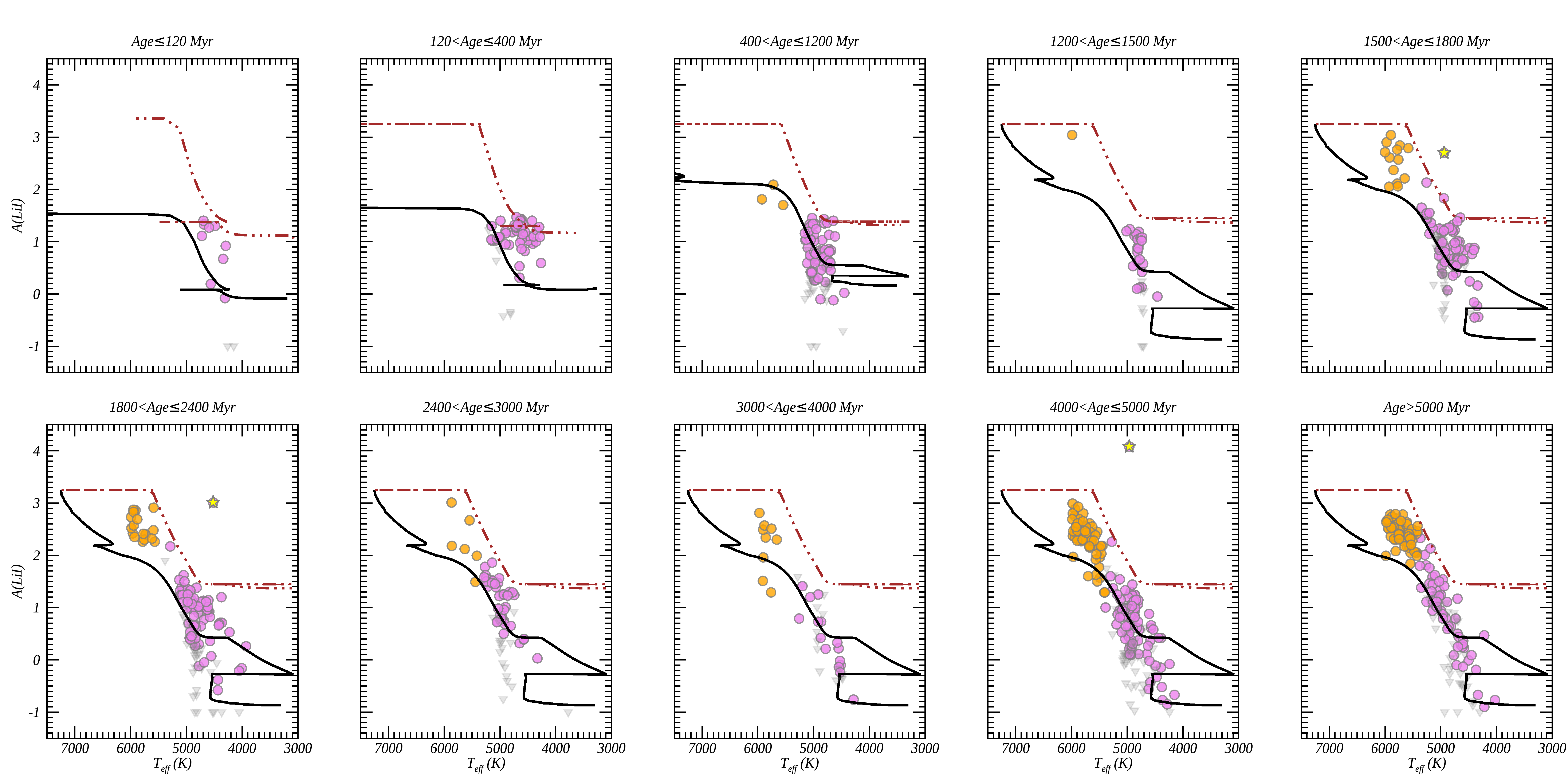}
\caption{A(Li) versus $T_\mathrm{eff}$ in open clusters, in different age bins. The curves are the model predictions from \citet{lagarde12} for the closest stellar masses, with standard mixing (brown dashed-line) and including rotation-induced mixing (black continuous line), and for solar metallicity.  Colours and symbols are as in  Fig.~\ref{fig:cluster1}.}
\label{fig:cluster:binned}
\end{figure*}

In Fig.~\ref{fig:cluster:binned}, we compare the predictions of the stellar evolutionary models with the cluster data, binned by age. This allows us to re-introduce the cluster member stars that were not included in Figs.~\ref{fig:cluster1} and \ref{fig:cluster2} because these clusters are too sparsely populated, and to have statistically significant samples in each age bin. We reach the same conclusions as in Sect.~\ref{subsection:Lifield}.
In the highest mass regime, corresponding to 120$\,\le\,$age$\,\le\,$400~Myr, the most important effect is rotation, which produces earlier and larger Li depletion than predicted by the classical model. 
In the clusters with 400~Myr$\,\le\,$age$\,\le\,$1200~Myr, the rotation-induced mixing  is needed to explain the behaviour of the lithium surface abundance in subgiant and giant stars,  while, as the age increases (hence the stellar mass decreases), the effect of the thermohaline mixing starts to play a role and can explain, together with the rotation-induced mixing, the further decrease in A(Li) for ages$\,>\,$1200~Myr. This effect is even stronger at ages$\,>\,$4000~Myr.

\section{Lithium-rich giant stars}
\label{sec:lithium:rich}

In this section we discuss Li-rich giant stars found both in open clusters and in the field. We recall the adopted definition of Li-rich giants: A(Li)$\,\ge\,$2.0~dex, 3800~K$\,\le T_\mathrm{eff}\le\,$5000~K, $\log g \le\,$3.5 or $\log(L/L_{\odot})\ge\,$1~dex and $\gamma \ge\,$0.98.
While red giant stars should have usually a lower Li surface abundance than in the previous evolutionary phases \citep[see, however,][and Magrini et al. (in prep.)]{kumar20b}, some of them present a clear Li overabundance with respect to the bulk.
A number of possible processes have been considered in the literature to explain the Li enrichment in these rare stars. Some works call for Li production thanks to deep internal transport processes  \citep[e.g.][]{1999ApJ...510..217S,2001A&A...375L...9P,2016A&A...585A.124C}. 
The Li enrichment has been also attributed to possible external pollution, such as planet engulfment or pollution by a binary companion  \citep[see, for instance, the case of Li-rich K giant][]{holanda20} or to magnetic activity \citep{goncalves20}. The ingestion of a planet or a companion brown dwarf can indeed contribute to increase the angular momentum of
the system, producing additional Li \citep[see, e.g.][]{alexander67,1999MNRAS.308.1133S,deni00,carlberg10,aguilera16,delgado16}.
The possible effects include an increase in the Li surface abundance, a change in the global metallicity and in rotational velocity \citep[see, e.g.][]{casey16}. Recently, several works show pollution of Be and Li in  white dwarfs, likely due to accretion of icy exomoons that formed around giant exoplanets or of other rocky bodies in exoplanetary systems \citep{klein21, doyle21, kaiser21}.
Other works, as \citet{jorissen20}, found that the binary frequency appears normal among the Li-rich giants, excluding a causal relationship between Li enrichment and binarity.  

\subsection{Lithium-rich giant stars and their evolutionary status}
The recent discovery of large samples of Li-rich giants indicates that they are not restricted just to the luminosity bump on the RGB for the low-mass stars, or its equivalent on the early-AGB for intermediate mass stars that ignite central He burning in non-degenerate conditions  \citep{cb00}, but they are also found along the RGB  and in the red clump \citep[see, e.g][]{alacala11, kumar11, lebzelter12, carlber16, smi18, deeppak19, c20, kumar20a, martell20, yan21}. 
The works of  \citet{kumar20a, yan21, singh21, DL21}, combining the results from asteroseismic and spectroscopic surveys, suggested that a high fraction of the Li-rich giants belong to the red clump central He-burning phase, confirming a previous idea presented by \citet{kumar11}. Similar results were obtained by \citet{casey19} where 80\% of their sample stars have likely helium-burning cores.

About 2\% of our giant stars in the field are Li-rich stars (considering as a reference sample giants in the same temperature range of the Li-rich ones), while the  percentage of Li-rich stars in open clusters is lower, 0.5\%.
These numbers agree with those found in several other surveys, which have reported the discovery of Li-rich giants in proportion of $\sim$1--2\% of their total sample \citep[see, e.g.][]{brown89, cb00, lebre06, kumar11, casey16, smi18, deeppak19, c20}.

In the field sample, we find 71 Li-rich giant stars, selected both with the criterion based on the gravity index $\gamma$ (21 stars) and with the surface gravity (50 stars). They are listed in Table~\ref{table:li:rich:field}, 
where we present their properties: their CNAME, GES field, SETUP, the stellar parameters, v$sin$i, A(Li) measured in the present work, and detection in previous works. 
Among the 50 stars selected through their $\log g$, we recovered 35 of the 40 Li-rich giants presented in \citet{casey16} and \citet{smi18}. The remaining 5 stars have  A(Li) slightly below 2.0~dex in {\sc idr6}, and thus they do not appear in our list of Li-rich giants with  A(Li)$\,\ge\,$2.0 dex.
The sample contains stars in the direction of the Bulge, 23 stars in the Corot fields (3 new discoveries) and one in the Kepler2 field, and stars in the field of several open clusters, but non-members.

\begin{figure*}[h!]
\centering
\includegraphics[scale=0.9]{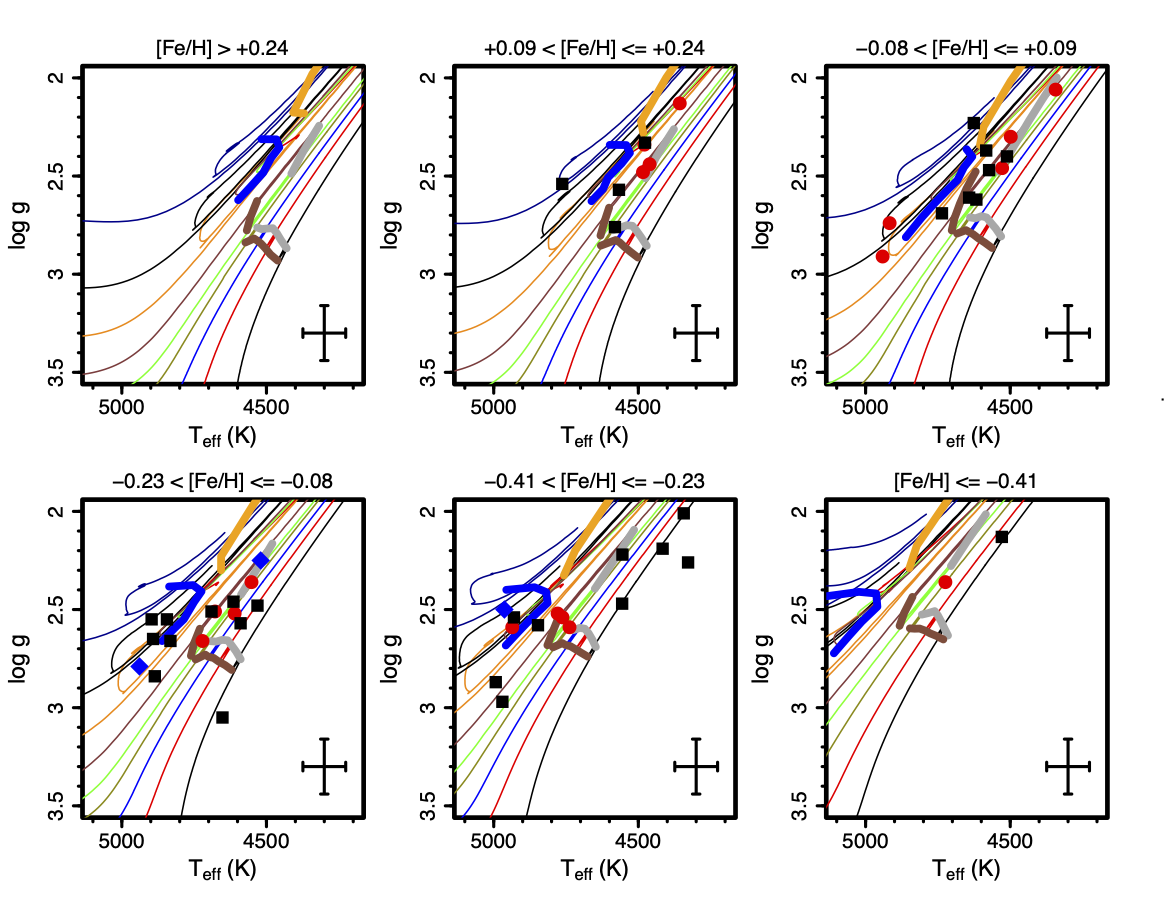}
\caption{Kiel diagram of the Li-rich *A(Li)$\,\ge\,$2.0~dex) giant stars compared with PARSEC evolutionary tracks \citep{bressan12, fu18} of masses 0.8, 1.0, 1.2, 1.4, 1.5, 1.7, 2.0, 2.4, and 3.0~$M_{\odot}$. From the top left to the bottom right panels, the models have [Fe/H] = +0.30, +0.18, 0.00, $-$0.15, $-$0.30, and $-$0.52 dex. The range of [Fe/H] of the stars is given at the top of each panel. Field stars with A(Li)$\,\ge\,$3.3~dex are shown as red circles, while stars with 2.0~dex$\,\le\,$A(Li)$\,<\,$3.3~dex are marked with black squares. The three Li-rich stars in open clusters are shown as blue diamonds. 
The beginning and the end of the RGB luminosity bump are marked as thick grey and brown lines, respectively.
The position of the clump of low-mass giants is shown as a thick blue line (from 0.8 to 1.9 $M_{\odot}$). 
The beginning of the early-AGB of intermediate-mass stars ($\ge\,$2.0~$M_{\odot}$) is highlighted as the thick orange line. Typical error bars are shown in the bottom right corner of the panels.
}
\label{fig:HR:fieldLirich}
\end{figure*}

In Fig.~\ref{fig:HR:fieldLirich} we show their location in the Kiel diagram, displaying only those with available $\log g$.  
In our data, there is no immediate correspondence of Li-rich stars with the position of the red clump, but instead we see 
a distribution around three main locations: the red giant branch luminosity bump, the core He-burning stages, or the early-asymptotic giant branch, as discussed for a sub-sample of these stars in \citet{smi18}.
As discussed by these authors, 
one of the CoRoT target, 101167637 (CNAME=19265193+0044004) is a confirmed He-core burning clump giant.  Their full characterisation with asteroseismology would thus be  necessary to reliably determine  the evolutionary status of each Li-rich star.

The three Li-rich giants detected as members of open clusters are listed in Table~\ref{tab:lirich:openclusters}, in which we give their properties: CNAME, the host cluster, the setup used to measure A(Li), stellar parameters, v$sin$i, and A(Li), and are shown in Fig.~\ref{fig:HR:fieldLirich}  with a different symbol.
They belong to Trumpler~5, Berkeley~21, and NGC~2158. The star in Trumpler~5 was first identified by \citet{monaco14} who attributed the Li enrichment to internal production, occurred at the red clump or in the immediately preceding phases.  The star in Berkeley~21 was discovered by \citet{hill99}, who measured its high Li content, relating it  to internal processes, but not discarding the possibility of an accretion from external sources. The Li-rich giant in NGC~2158 is, to our knowledge, a new detection. Two of the three Li-rich stars are  located close to the RC, and one of them at the beginning of the RGB luminosity bump.  They do not have an enhanced rotational velocity, as would be expected if the Li-enrichment were related, e.g., to planet engulfment \citep[see, e.g.][and the discussion in Sec~\ref{sec:rotation}]{privitera16}.

\begin{table*}
\caption{Red giant Li-rich stars in the field. The full table is available online at CDS.}
\label{table:li:rich:field}
\tiny
\begin{tabular}{lllllllll}
\hline\hline
 \multicolumn{1}{c}{CNAME} &
   \multicolumn{1}{c}{Field} &
  \multicolumn{1}{c}{SETUP} &
  \multicolumn{1}{c}{$T_\mathrm{eff}$} &
  \multicolumn{1}{c}{log~g} &
  \multicolumn{1}{c}{[Fe/H]} &
  \multicolumn{1}{c}{v$sin$i} &
  \multicolumn{1}{c}{A(Li)}  &
  \multicolumn{1}{c}{Other detection} \\
\hline
  06490710-2359450 & Be75 (non member) & HR15N & 4993$\pm$60 & 2.9$\pm$0.2 & -0.29$\pm$0.05 & $\leq$7.0 & 2.19$\pm$0.08 &  -- \\
  18182698-3242584 & Bulge & U580 & 4340$\pm$30 & 2.06$\pm$0.06 & 0.06$\pm$0.05 & 9.0   & 3.68$\pm$0.05 & \citet{smi18}\\
  18181062-3246291 & Bulge & U580 & 4580$\pm$30 & 2.37$\pm$0.05 & 0.06$\pm$0.05 & 8.0   & 2.01$\pm$0.05 & \citet{smi18}\\
  18033785-3009201 & Bulge & U580 & 4480$\pm$30 & 2.48$\pm$0.05 & 0.13$\pm$0.05  & 8.0 & 3.61$\pm$0.04 & \citet{smi18}\\
\hline\end{tabular}
\end{table*}

\begin{table*}
\caption{Red giant Li-rich stars in open clusters. }
\label{tab:lirich:openclusters}
\tiny
\begin{tabular}{lllllllll}
\hline\hline
  \multicolumn{1}{c}{CNAME} &
  \multicolumn{1}{c}{Clusters} &
  \multicolumn{1}{c}{SETUP} &
  \multicolumn{1}{c}{$T_\mathrm{eff}$} &
  \multicolumn{1}{c}{log~g} &
  \multicolumn{1}{c}{[Fe/H]} &
  \multicolumn{1}{c}{v$sin$i} &
  \multicolumn{1}{c}{A(Li)}  &
   \multicolumn{1}{c }{Other detection} \\
\hline
  06364020+0929478 & Trumpler5 & U580  & 4960$\pm$30 & 2.50$\pm$0.05 & -0.37$\pm$0.04 & $\leq$7.0 & 4.08$\pm$0.05 & \citet{monaco14}\\
  05514200+2148497 & Be21 & U580 & 4520$\pm$30 & 2.25$\pm$0.05 & -0.18$\pm$0.04 & $\leq$7.0  & 3.01$\pm$0.06 & \citet{hill99}\\
  06072443+2400524 & NGC2158 & HR15N  & 4940$\pm$60 & 2.8$\pm$0.2 & -0.17$\pm$0.06 & $\leq$7.0  & 2.70$\pm$0.09 & --\\
\hline\end{tabular}
\end{table*}

\subsection{Li abundance and rotational velocity}
\label{sec:rotation}

The relation between rotation and lithium abundance in evolved stars is  not yet completely established \citep[see, e.g.][]{walle94, demedeiros97, demedeiros00,  delaverny03, mallik03}.
As discussed in \citet{smi18}, fast rotation during the giant phase cannot be explained by single star evolution and might be related to planet engulfment \citep{alexander67, carlberg09, carlberg10, casey16, privitera16, anthony20}.    
 Following \citet{privitera16}, the planet engulfment might have an important effect of the Li abundances, even for a limited portion of the giant life.  However, the effect is difficult to disentangle from other processes that can modify the Li abundance, since Li is  a fragile element, easily destroyed, and subject to other mechanisms of production during the red giant phase. 
\citet{delgado16} looked for Li-rich giants in a sample of clusters where planets have been searched,  deriving A(Li) abundances in 12 open clusters. They studied the relationship between v$sin$i and A(Li), finding that the giant stars with higher A(Li) have higher rotation velocities than the Li-depleted stars. However, they found also that the relation might reflect the different evolutionary status of their sample stars, with the hottest stars having higher rotation rates.

\begin{figure}[h!]
\centering
\includegraphics[width=\hsize]{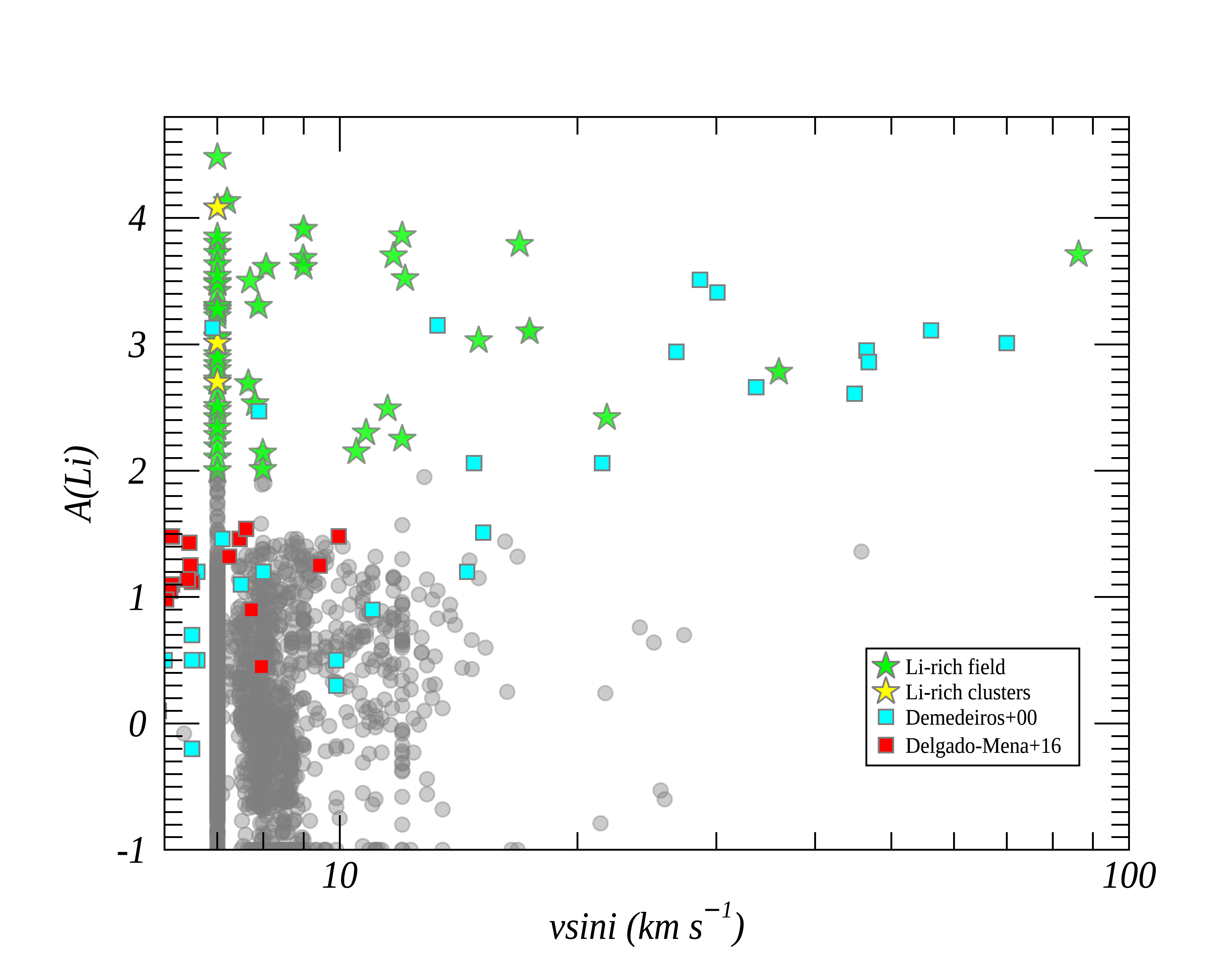}
\caption{A(Li) as a function of the projected rotational velocity, v$sin$i, in our sample of giant stars with 3800~K$\,\le T_\mathrm{eff}\leq\,$5000~K and $\log g \le\,$3.5: grey circles are the giant stars in open clusters and in the field with A(Li)$\,<\,$2.0~dex, while light yellow and green stars represent A(Li)$\,\ge\,$2~dex in giants  in open clusters and in the field, respectively; red squares are the giant stars observed by \citet{delgado16}, while cyan squares are those observed by \citet{demedeiros00}.   }
\label{fig:li:vrot:all}
\end{figure}

For our sample of giant stars, both in field and in clusters,  we seek possible correlations between A(Li) and  the projected rotational velocities.
In a conservative way, we consider  fast-rotating giant stars those with v$sin$i$\,>$ 10~km~s$^{-1}$. In our sample, the instrumental limit prevents us to measure v$sin$i$\,\le\,$7~km~s$^{-1}$, thus for them we can provide only an upper limit.
The results are shown in Fig.~\ref{fig:li:vrot:all}, where we plot A(Li) as a function of the projected rotational velocity for giants with 3800~K$\,\le T_\mathrm{eff}\leq\,$5000~K and $\log g \le\,$3.5, both Li-rich (A(Li)$\,\ge\,$2) and normal ones. The samples of  giant stars  observed by \citet{demedeiros00} and by \citet{delgado16} are  over-plotted for comparison.  
We notice a trend of increasing A(Li) with increasing  v$sin$i. However,  for 10~km~s$^{-1}\le\,$v$sin$i$\,\le\,$30~km~s$^{-1}$ we observe both stars with A(Li) above and below 2.0~dex. 
Many Li-rich stars have low v$sin$i, including the three Li-rich stars in open clusters, indicating that the preferential way to produce Li enrichment in giant stars is related to some specific phases of stellar evolution, as shown in Fig.~\ref{fig:HR:fieldLirich}, and that the correlation with the projected rotational velocity indicates that lithium enrichment by engulfment is an occasional effect. 
 

\section{Summary and conclusions}
\label{section:SummaryConclusions}

We exploit a sample of giant stars with Li measurements in Gaia-ESO {\sc idr6} to investigate the evolution of A(Li) from the MSTO to the giant phase. 
We combine the Gaia-ESO data with Gaia {\sc edr3} to obtain the distances and stellar luminosities. We compare our lithium abundances with literature values, finding a good agreement. 

We select  MS, sub-giant, and giant stars which are member stars of open clusters and field stars. We study the general trends of lithium abundances after the MS.  
Since stellar masses play a fundamental role during the post-main sequence evolution, we  select samples  of  stars with reliable measurements of stellar masses: member stars of open clusters (age$\,>\,$120 Myr) and field  stars with masses derived from isochrone fitting. Our data probe, with a homogeneous analysis, lithium abundances and stellar parameters for stars having a wide range of stellar masses, covered thanks to the sample of young and intermediate-age open clusters (with TO masses from 1.1 to 4.5~$M_{\odot}$) not usually available in surveys that account only for field stars. 

We compare our results with the set of stellar models of \citet{lagarde12}, in which the effect of rotation-induced mixing and thermohaline instability are included.
The comparison between our data and model results confirm the strong impact of the rotation-induced mixing, already in massive stars.  The lower mass giant stars, both in clusters and in the field, provide also support to the necessity of a mixing process in advanced phases of stellar evolution, which might be  thermohaline mixing.
We confirm the agreement between data and models with rotation-induced and thermohaline mixing  in the whole mass and metallicity ranges. 

We discuss the properties of our Li-rich sample of stars, including both field stars and a few members of open clusters. They are distributed around three main locations in the Kiel diagram: the red giant branch luminosity bump, the core-He burning stages, or the early-asymptotic giant branch. Their full characterisation with asteroseismology would be  necessary to establish a unique link with  the evolutionary status of each stars.
Finally, we investigate possible effects of the residual stellar rotation, after the MS,  during the giant phase. We find few stars with v$sin$i$\,>\,$10~km~s$^{-1}$, and their Li abundance is 
in line with the other stars in the same evolutionary state. We do not find any conclusive correlation between Li-rich stars and projected rotational velocity.  

\begin{acknowledgements}
We thank the referee for her/his careful reading of the paper, and for comments, which improved the quality of the work and its presentation. Based on data products from observations made with ESO Telescopes at the La Silla Paranal Observatory under programme ID 188.B-3002. These data products have been processed by the Cambridge Astronomy Survey Unit (CASU) at the Institute of Astronomy, University of Cambridge, and by the FLAMES/UVES reduction team at INAF/Osservatorio Astrofisico di Arcetri. These data have been obtained from the Gaia-ESO Survey Data Archive, prepared and hosted by the Wide Field Astronomy Unit, Institute for Astronomy, University of Edinburgh, which is funded by the UK Science and Technology Facilities Council.
This work was partly supported by the European Union FP7 programme through ERC grant number 320360 and by the Leverhulme Trust through grant RPG-2012-541. We acknowledge the support from INAF and Ministero dell' Istruzione, dell' Universit\`a' e della Ricerca (MIUR) in the form of the grant "Premiale VLT 2012". The results presented here benefit from discussions held during the Gaia-ESO workshops and conferences supported by the ESF (European Science Foundation) through the GREAT Research Network Programme.
LM, GC, SR, MVdS acknowledge the funding from MIUR Premiale 2016: MITiC. MVdS, LM and AV thank the WEAVE-Italia consortium. LM acknowledge the funding from the INAF PRIN-SKA 2017 program 1.05.01.88.04. NL acknowledges the "Programme National de Physique Stellaire" (PNPS) and the "Programme National Cosmology et Galaxies (PNCG)" of CNRS/INSU co-funded by CEA and CNES. 
CVV and LM thank the COST Action CA18104: MW-Gaia. LS acknowledges financial support from the Australian Research Council (discovery Project 170100521) and from the Australian Research Council Centre of Excellence for All Sky Astrophysics in 3 Dimensions (ASTRO 3D), through project number CE170100013.  
TB acknowledges financial support by grant No. 2018-04857 from the Swedish Research Council.
EDM acknowledges the support from Funda\c{c}\~ao para a Ci\^encia e a Tecnologia (FCT) through the
research grants UIDB/04434/2020 and UIDP/04434/2020 and by the Investigador FCT contract IF/00849/2015.
FJE acknowledges financial support from the Spanish MINECO/FEDER through the grant AYA2017-84089 and MDM-2017-0737 at Centro de Astrobiología (CSIC-INTA), Unidad de Excelencia María de Maeztu, and from the European Union’s Horizon 2020 research and innovation programme under Grant Agreement no. 824064 through the ESCAPE - The European Science Cluster of Astronomy \& Particle Physics ESFRI Research Infrastructures project. ASB acknowledges the financial support of the STFC. E.D.M. acknowledges the support from Funda\c{c}\~ao para a Ci\^encia e a Tecnologia (FCT) through national funds and from FEDER through COMPETE2020 by the grant UIDB/04434/2020 \& UIDP/04434/2020 and by the Investigador FCT contract IF/00849/2015.

\end{acknowledgements}

\bibliographystyle{aa}
\bibliography{Bibliography}

\begin{thebibliography}{155}
\expandafter\ifx\csname natexlab\endcsname\relax\def\natexlab#1{#1}\fi

\bibitem[{{Aguilera-G{\'o}mez} {et~al.}(2016){Aguilera-G{\'o}mez},
  {Chanam{\'e}}, {Pinsonneault}, \& {Carlberg}}]{aguilera16}
{Aguilera-G{\'o}mez}, C., {Chanam{\'e}}, J., {Pinsonneault}, M.~H., \&
  {Carlberg}, J.~K. 2016, \apjl, 833, L24

\bibitem[{{Alcal{\'a}} {et~al.}(2011){Alcal{\'a}}, {Biazzo}, {Covino},
  {Frasca}, \& {Bedin}}]{alacala11}
{Alcal{\'a}}, J.~M., {Biazzo}, K., {Covino}, E., {Frasca}, A., \& {Bedin},
  L.~R. 2011, \aap, 531, L12

\bibitem[{{Alexander}(1967)}]{alexander67}
{Alexander}, J.~B. 1967, The Observatory, 87, 238

\bibitem[{{Alonso} {et~al.}(1999){Alonso}, {Arribas}, \&
  {Mart{\'\i}nez-Roger}}]{alonso99}
{Alonso}, A., {Arribas}, S., \& {Mart{\'\i}nez-Roger}, C. 1999, \aaps, 140, 261

\bibitem[{{Angelou} {et~al.}(2015){Angelou}, {D'Orazi}, {Constantino},
  {Church}, {Stancliffe}, \& {Lattanzio}}]{2015MNRAS.450.2423A}
{Angelou}, G.~C., {D'Orazi}, V., {Constantino}, T.~N., {et~al.} 2015, \mnras,
  450, 2423

\bibitem[{{Anthony-Twarog} {et~al.}(2018){Anthony-Twarog}, {Deliyannis},
  {Harmer}, {Lee-Brown}, {Steinhauer}, {Sun}, \& {Twarog}}]{anthony18}
{Anthony-Twarog}, B.~J., {Deliyannis}, C.~P., {Harmer}, D., {et~al.} 2018, \aj,
  156, 37

\bibitem[{{Anthony-Twarog} {et~al.}(2013){Anthony-Twarog}, {Deliyannis},
  {Rich}, \& {Twarog}}]{at12}
{Anthony-Twarog}, B.~J., {Deliyannis}, C.~P., {Rich}, E., \& {Twarog}, B.~A.
  2013, \apjl, 767, L19

\bibitem[{{Anthony-Twarog} {et~al.}(2020){Anthony-Twarog}, {Deliyannis}, \&
  {Twarog}}]{anthony20}
{Anthony-Twarog}, B.~J., {Deliyannis}, C.~P., \& {Twarog}, B.~A. 2020, \aj,
  160, 75

\bibitem[{{Anthony-Twarog} {et~al.}(2009){Anthony-Twarog}, {Deliyannis},
  {Twarog}, {Croxall}, \& {Cummings}}]{2009AJ....138.1171A}
{Anthony-Twarog}, B.~J., {Deliyannis}, C.~P., {Twarog}, B.~A., {Croxall},
  K.~V., \& {Cummings}, J.~D. 2009, \aj, 138, 1171

\bibitem[{{Asplund} {et~al.}(2009){Asplund}, {Grevesse}, {Sauval}, \&
  {Scott}}]{asplund09}
{Asplund}, M., {Grevesse}, N., {Sauval}, A.~J., \& {Scott}, P. 2009, \araa, 47,
  481

\bibitem[{{Bailer-Jones} {et~al.}(2020){Bailer-Jones}, {Rybizki}, {Fouesneau},
  {Demleitner}, \& {Andrae}}]{bailerjones20}
{Bailer-Jones}, C.~A.~L., {Rybizki}, J., {Fouesneau}, M., {Demleitner}, M., \&
  {Andrae}, R. 2020, arXiv e-prints, arXiv:2012.05220

\bibitem[{{Balachandran}(1990)}]{Balachandran90}
{Balachandran}, S. 1990, \apj, 354, 310

\bibitem[{{Balachandran}(1995)}]{1995MmSAI..66..387B}
{Balachandran}, S. 1995, \memsai, 66, 387

\bibitem[{{Baraffe} {et~al.}(2017){Baraffe}, {Pratt}, {Goffrey}, {Constantino},
  {Folini}, {Popov}, {Walder}, \& {Viallet}}]{2017ApJ...845L...6B}
{Baraffe}, I., {Pratt}, J., {Goffrey}, T., {et~al.} 2017, \apjl, 845, L6

\bibitem[{{Bensby} \& {Lind}(2018)}]{bensby18}
{Bensby}, T. \& {Lind}, K. 2018, \aap, 615, A151

\bibitem[{{Bovy}(2016)}]{bovy16}
{Bovy}, J. 2016, \apj, 817, 49

\bibitem[{{Bravi} {et~al.}(2018){Bravi}, {Zari}, {Sacco}, {Randich},
  {Jeffries}, {Jackson}, {Franciosini}, {Moraux}, {L{\'o}pez-Santiago},
  {Pancino}, {Spina}, {Wright}, {Jim{\'e}nez-Esteban}, {Klutsch},
  {Roccatagliata}, {Gilmore}, {Bragaglia}, {Flaccomio}, {Francois}, {Koposov},
  {Bayo}, {Carraro}, {Costado}, {Damiani}, {Frasca}, {Hourihane}, {Jofr{\'e}},
  {Lardo}, {Lewis}, {Magrini}, {Morbidelli}, {Prisinzano}, {Sousa}, {Worley},
  \& {Zaggia}}]{bravi18}
{Bravi}, L., {Zari}, E., {Sacco}, G.~G., {et~al.} 2018, \aap, 615, A37

\bibitem[{{Bressan} {et~al.}(2012){Bressan}, {Marigo}, {Girardi}, {Salasnich},
  {Dal Cero}, {Rubele}, \& {Nanni}}]{bressan12}
{Bressan}, A., {Marigo}, P., {Girardi}, L., {et~al.} 2012, \mnras, 427, 127

\bibitem[{{Brown} {et~al.}(1989){Brown}, {Sneden}, {Lambert}, \&
  {Dutchover}}]{brown89}
{Brown}, J.~A., {Sneden}, C., {Lambert}, D.~L., \& {Dutchover}, Edward, J.
  1989, \apjs, 71, 293

\bibitem[{{Buder} {et~al.}(2020){Buder}, {Sharma}, {Kos}, {Amarsi},
  {Nordlander}, {Lind}, {Martell}, {Asplund}, {Bland-Hawthorn}, {Casey}, {De
  Silva}, {D'Orazi}, {Freeman}, {Hayden}, {Lewis}, {Lin}, {Schlesinger},
  {Simpson}, {Stello}, {Zucker}, {Zwitter}, {Beeson}, {Buck}, {Casagrande},
  {Clark}, {Cotar}, {Da Costa}, {de Grijs}, {Feuillet}, {Horner}, {Khanna},
  {Kafle}, {Liu}, {Montet}, {Nandakumar}, {Nataf}, {Ness}, {Spina}, {Traven},
  {Trepper-Garcia}, {Ting}, {Vogrincic}, {Wittenmyer}, {Zerjal}, \& {the GALAH
  collaboration}}]{buder20}
{Buder}, S., {Sharma}, S., {Kos}, J., {et~al.} 2020, arXiv e-prints,
  arXiv:2011.02505

\bibitem[{{Cantat-Gaudin} {et~al.}(2020){Cantat-Gaudin}, {Anders},
  {Castro-Ginard}, {Jordi}, {Romero-G{\'o}mez}, {Soubiran}, {Casamiquela},
  {Tarricq}, {Moitinho}, {Vallenari}, {Bragaglia}, {Krone-Martins}, \&
  {Kounkel}}]{CG20}
{Cantat-Gaudin}, T., {Anders}, F., {Castro-Ginard}, A., {et~al.} 2020, \aap,
  640, A1

\bibitem[{{Canto Martins} {et~al.}(2011){Canto Martins}, {L{\`e}bre},
  {Palacios}, {de Laverny}, {Richard}, {Melo}, {Do Nascimento}, \& {de
  Medeiros}}]{canto11}
{Canto Martins}, B.~L., {L{\`e}bre}, A., {Palacios}, A., {et~al.} 2011, \aap,
  527, A94

\bibitem[{{Carlberg} {et~al.}(2016){Carlberg}, {Cunha}, \& {Smith}}]{carlber16}
{Carlberg}, J.~K., {Cunha}, K., \& {Smith}, V.~V. 2016, \apj, 827, 129

\bibitem[{{Carlberg} {et~al.}(2009){Carlberg}, {Majewski}, \&
  {Arras}}]{carlberg09}
{Carlberg}, J.~K., {Majewski}, S.~R., \& {Arras}, P. 2009, \apj, 700, 832

\bibitem[{{Carlberg} {et~al.}(2015){Carlberg}, {Smith}, {Cunha}, {Majewski},
  {M{\'e}sz{\'a}ros}, {Shetrone}, {Allende Prieto}, {Bizyaev}, {Stassun},
  {Fleming}, {Zasowski}, {Hearty}, {Nidever}, {Schneider}, {Holtzman}, \&
  {Frinchaboy}}]{carlberg15}
{Carlberg}, J.~K., {Smith}, V.~V., {Cunha}, K., {et~al.} 2015, \apj, 802, 7

\bibitem[{{Carlberg} {et~al.}(2010){Carlberg}, {Smith}, {Cunha}, {Majewski}, \&
  {Rood}}]{carlberg10}
{Carlberg}, J.~K., {Smith}, V.~V., {Cunha}, K., {Majewski}, S.~R., \& {Rood},
  R.~T. 2010, \apjl, 723, L103

\bibitem[{{Carrera} \& {Mart{\'\i}nez-V{\'a}zquez}(2013)}]{carrera13}
{Carrera}, R. \& {Mart{\'\i}nez-V{\'a}zquez}, C.~E. 2013, \aap, 560, A5

\bibitem[{{Casagrande} \& {VandenBerg}(2018)}]{casagrande18}
{Casagrande}, L. \& {VandenBerg}, D.~A. 2018, \mnras, 479, L102

\bibitem[{{Casey} {et~al.}(2019){Casey}, {Ho}, {Ness}, {Hogg}, {Rix},
  {Angelou}, {Hekker}, {Tout}, {Lattanzio}, {Karakas}, {Woods}, {Price-Whelan},
  \& {Schlaufman}}]{casey19}
{Casey}, A.~R., {Ho}, A. Y.~Q., {Ness}, M., {et~al.} 2019, \apj, 880, 125

\bibitem[{{Casey} {et~al.}(2016){Casey}, {Ruchti}, {Masseron}, {Randich},
  {Gilmore}, {Lind}, {Kennedy}, {Koposov}, {Hourihane}, {Franciosini}, {Lewis},
  {Magrini}, {Morbidelli}, {Sacco}, {Worley}, {Feltzing}, {Jeffries},
  {Vallenari}, {Bensby}, {Bragaglia}, {Flaccomio}, {Francois}, {Korn},
  {Lanzafame}, {Pancino}, {Recio-Blanco}, {Smiljanic}, {Carraro}, {Costado},
  {Damiani}, {Donati}, {Frasca}, {Jofr{\'e}}, {Lardo}, {de Laverny}, {Monaco},
  {Prisinzano}, {Sbordone}, {Sousa}, {Tautvai{\v{s}}ien{\.{e}}}, {Zaggia},
  {Zwitter}, {Delgado Mena}, {Chorniy}, {Martell}, {Silva Aguirre}, {Miglio},
  {Chiappini}, {Montalban}, {Morel}, \& {Valentini}}]{casey16}
{Casey}, A.~R., {Ruchti}, G., {Masseron}, T., {et~al.} 2016, \mnras, 461, 3336

\bibitem[{{Cassisi} {et~al.}(2016){Cassisi}, {Salaris}, \&
  {Pietrinferni}}]{2016A&A...585A.124C}
{Cassisi}, S., {Salaris}, M., \& {Pietrinferni}, A. 2016, \aap, 585, A124

\bibitem[{{Castro} {et~al.}(2016){Castro}, {Duarte}, {Pace}, \& {do
  Nascimento}}]{Castroetal2016}
{Castro}, M., {Duarte}, T., {Pace}, G., \& {do Nascimento}, J.~D. 2016, \aap,
  590, A94

\bibitem[{{Ceillier} {et~al.}(2013){Ceillier}, {Eggenberger}, {Garc{\'\i}a}, \&
  {Mathis}}]{2013A&A...555A..54C}
{Ceillier}, T., {Eggenberger}, P., {Garc{\'\i}a}, R.~A., \& {Mathis}, S. 2013,
  \aap, 555, A54

\bibitem[{{Charbonneau} \& {Michaud}(1990)}]{CharbonneauMichaud1990}
{Charbonneau}, P. \& {Michaud}, G. 1990, \apj, 352, 681

\bibitem[{{Charbonnel} \& {Balachandran}(2000)}]{cb00}
{Charbonnel}, C. \& {Balachandran}, S.~C. 2000, \aap, 359, 563

\bibitem[{{Charbonnel} {et~al.}(1998){Charbonnel}, {Brown}, \&
  {Wallerstein}}]{1998A&A...332..204C}
{Charbonnel}, C., {Brown}, J.~A., \& {Wallerstein}, G. 1998, \aap, 332, 204

\bibitem[{{Charbonnel} \& {Lagarde}(2010)}]{CL10}
{Charbonnel}, C. \& {Lagarde}, N. 2010, \aap, 522, A10

\bibitem[{{Charbonnel} {et~al.}(2020){Charbonnel}, {Lagarde}, {Jasniewicz},
  {North}, {Shetrone}, {Krugler Hollek}, {Smith}, {Smiljanic}, {Palacios}, \&
  {Ottoni}}]{c20}
{Charbonnel}, C., {Lagarde}, N., {Jasniewicz}, G., {et~al.} 2020, \aap, 633,
  A34

\bibitem[{{Charbonnel} \& {Zahn}(2007)}]{cczahn07}
{Charbonnel}, C. \& {Zahn}, J.~P. 2007, \aap, 467, L15

\bibitem[{{Christensen-Dalsgaard} {et~al.}(2011){Christensen-Dalsgaard},
  {Monteiro}, {Rempel}, \& {Thompson}}]{CD11}
{Christensen-Dalsgaard}, J., {Monteiro}, M.~J.~P.~F.~G., {Rempel}, M., \&
  {Thompson}, M.~J. 2011, \mnras, 414, 1158

\bibitem[{{Coc} {et~al.}(2004){Coc}, {Vangioni-Flam}, {Descouvemont},
  {Adahchour}, \& {Angulo}}]{2004ApJ...600..544C}
{Coc}, A., {Vangioni-Flam}, E., {Descouvemont}, P., {Adahchour}, A., \&
  {Angulo}, C. 2004, \apj, 600, 544

\bibitem[{{Damiani} {et~al.}(2014){Damiani}, {Prisinzano}, {Micela}, {Randich},
  {Gilmore}, {Drew}, {Jeffries}, {Fr{\'e}mat}, {Alfaro}, {Bensby}, {Bragaglia},
  {Flaccomio}, {Lanzafame}, {Pancino}, {Recio-Blanco}, {Sacco}, {Smiljanic},
  {Jackson}, {de Laverny}, {Morbidelli}, {Worley}, {Hourihane}, {Costado},
  {Jofr{\'e}}, {Lind}, \& {Maiorca}}]{damiani14}
{Damiani}, F., {Prisinzano}, L., {Micela}, G., {et~al.} 2014, \aap, 566, A50

\bibitem[{{de Laverny} {et~al.}(2003){de Laverny}, {do Nascimento},
  {L{\`e}bre}, \& {De Medeiros}}]{delaverny03}
{de Laverny}, P., {do Nascimento}, J.~D., J., {L{\`e}bre}, A., \& {De
  Medeiros}, J.~R. 2003, \aap, 410, 937

\bibitem[{{de Laverny} {et~al.}(2012){de Laverny}, {Recio-Blanco}, {Worley}, \&
  {Plez}}]{delaverny12}
{de Laverny}, P., {Recio-Blanco}, A., {Worley}, C.~C., \& {Plez}, B. 2012,
  \aap, 544, A126

\bibitem[{{de Medeiros} {et~al.}(1997){de Medeiros}, {Do Nascimento}, \&
  {Mayor}}]{demedeiros97}
{de Medeiros}, J.~R., {Do Nascimento}, J.~D., J., \& {Mayor}, M. 1997, \aap,
  317, 701

\bibitem[{{De Medeiros} {et~al.}(2000){De Medeiros}, {do Nascimento},
  {Sankarankutty}, {Costa}, \& {Maia}}]{demedeiros00}
{De Medeiros}, J.~R., {do Nascimento}, J.~D., J., {Sankarankutty}, S., {Costa},
  J.~M., \& {Maia}, M.~R.~G. 2000, \aap, 363, 239

\bibitem[{{De Silva} {et~al.}(2006){De Silva}, {Sneden}, {Paulson}, {Asplund},
  {Bland-Hawthorn}, {Bessell}, \& {Freeman}}]{desilva06}
{De Silva}, G.~M., {Sneden}, C., {Paulson}, D.~B., {et~al.} 2006, \aj, 131, 455

\bibitem[{{Deal} {et~al.}(2021){Deal}, {Richard}, \&
  {Vauclair}}]{2021A&A...646A.160D}
{Deal}, M., {Richard}, O., \& {Vauclair}, S. 2021, \aap, 646, A160

\bibitem[{{Deepak} \& {Lambert}(2021)}]{DL21}
{Deepak} \& {Lambert}, D.~L. 2021, arXiv e-prints, arXiv:2104.11741

\bibitem[{{Deepak} {et~al.}(2020){Deepak}, {Lambert}, \& {Reddy}}]{deepak20}
{Deepak}, {Lambert}, D.~L., \& {Reddy}, B.~E. 2020, \mnras, 494, 1348

\bibitem[{{Deepak} \& {Reddy}(2019)}]{deeppak19}
{Deepak} \& {Reddy}, B.~E. 2019, \mnras, 484, 2000

\bibitem[{{Delgado Mena} {et~al.}(2016){Delgado Mena}, {Tsantaki}, {Sousa},
  {Kunitomo}, {Adibekyan}, {Zaworska}, {Santos}, {Israelian}, \&
  {Lovis}}]{delgado16}
{Delgado Mena}, E., {Tsantaki}, M., {Sousa}, S.~G., {et~al.} 2016, \aap, 587,
  A66

\bibitem[{{Deliyannis} {et~al.}(2019){Deliyannis}, {Anthony-Twarog},
  {Lee-Brown}, \& {Twarog}}]{Deliyannis19}
{Deliyannis}, C.~P., {Anthony-Twarog}, B.~J., {Lee-Brown}, D.~B., \& {Twarog},
  B.~A. 2019, \aj, 158, 163

\bibitem[{{Deliyannis} {et~al.}(2000){Deliyannis}, {Pinsonneault}, \&
  {Charbonnel}}]{2000IAUS..198...61D}
{Deliyannis}, C.~P., {Pinsonneault}, M.~H., \& {Charbonnel}, C. 2000, in IAU
  Symposium, Vol. 198, The Light Elements and their Evolution, ed. L.~{da
  Silva}, R.~{de Medeiros}, \& M.~{Spite}, 61

\bibitem[{{Denissenkov} {et~al.}(2009){Denissenkov}, {Pinsonneault}, \&
  {MacGregor}}]{2009ApJ...696.1823D}
{Denissenkov}, P.~A., {Pinsonneault}, M., \& {MacGregor}, K.~B. 2009, \apj,
  696, 1823

\bibitem[{{Denissenkov} \& {Tout}(2003)}]{denissenkov03}
{Denissenkov}, P.~A. \& {Tout}, C.~A. 2003, \mnras, 340, 722

\bibitem[{{Denissenkov} \& {Weiss}(2000)}]{deni00}
{Denissenkov}, P.~A. \& {Weiss}, A. 2000, \aap, 358, L49

\bibitem[{{Doyle} {et~al.}(2021){Doyle}, {Desch}, \& {Young}}]{doyle21}
{Doyle}, A.~E., {Desch}, S.~J., \& {Young}, E.~D. 2021, \apjl, 907, L35

\bibitem[{{Dumont} {et~al.}(2021){Dumont}, {Palacios}, {Charbonnel}, {Richard},
  {Amard}, {Augustson}, \& {Mathis}}]{2021A&A...646A..48D}
{Dumont}, T., {Palacios}, A., {Charbonnel}, C., {et~al.} 2021, \aap, 646, A48

\bibitem[{{Eggenberger} {et~al.}(2019){Eggenberger}, {Deheuvels}, {Miglio},
  {Ekstr{\"o}m}, {Georgy}, {Meynet}, {Lagarde}, {Salmon}, {Buldgen},
  {Montalb{\'a}n}, {Spada}, \& {Ballot}}]{2019A&A...621A..66E}
{Eggenberger}, P., {Deheuvels}, S., {Miglio}, A., {et~al.} 2019, \aap, 621, A66

\bibitem[{{Eggenberger} {et~al.}(2012){Eggenberger}, {Haemmerl{\'e}}, {Meynet},
  \& {Maeder}}]{2012A&A...539A..70E}
{Eggenberger}, P., {Haemmerl{\'e}}, L., {Meynet}, G., \& {Maeder}, A. 2012,
  \aap, 539, A70

\bibitem[{{Eggenberger} {et~al.}(2017){Eggenberger}, {Lagarde}, {Miglio},
  {Montalb{\'a}n}, {Ekstr{\"o}m}, {Georgy}, {Meynet}, {Salmon}, {Ceillier},
  {Garc{\'\i}a}, {Mathis}, {Deheuvels}, {Maeder}, {den Hartogh}, \&
  {Hirschi}}]{2017A&A...599A..18E}
{Eggenberger}, P., {Lagarde}, N., {Miglio}, A., {et~al.} 2017, \aap, 599, A18

\bibitem[{{Fran{\c{c}}ois} {et~al.}(2013){Fran{\c{c}}ois}, {Pasquini},
  {Biazzo}, {Bonifacio}, \& {Palsa}}]{francois13}
{Fran{\c{c}}ois}, P., {Pasquini}, L., {Biazzo}, K., {Bonifacio}, P., \&
  {Palsa}, R. 2013, \aap, 552, A136

\bibitem[{{Franciosini} {et~al.}(2018){Franciosini}, {Sacco}, {Jeffries},
  {Damiani}, {Roccatagliata}, {Fedele}, \& {Randich}}]{francio18}
{Franciosini}, E., {Sacco}, G.~G., {Jeffries}, R.~D., {et~al.} 2018, \aap, 616,
  L12

\bibitem[{{Fu} {et~al.}(2018){Fu}, {Bressan}, {Marigo}, {Girardi},
  {Montalb{\'a}n}, {Chen}, \& {Nanni}}]{fu18}
{Fu}, X., {Bressan}, A., {Marigo}, P., {et~al.} 2018, \mnras, 476, 496

\bibitem[{{Gaia Collaboration} {et~al.}(2021){Gaia Collaboration}, {Brown},
  {Vallenari}, {Prusti}, {de Bruijne}, {Babusiaux}, {Biermann},
  {et~al.}}]{gaia20}
{Gaia Collaboration}, {Brown}, A.~G.~A., {Vallenari}, A., {et~al.} 2021, \aap,
  649, A1

\bibitem[{{Galli} \& {Palla}(2013)}]{GP13}
{Galli}, D. \& {Palla}, F. 2013, \araa, 51, 163

\bibitem[{{Garaud}(2021)}]{2021arXiv210308072G}
{Garaud}, P. 2021, arXiv e-prints, arXiv:2103.08072

\bibitem[{{Garaud} \& {Kulenthirarajah}(2016)}]{2016ApJ...821...49G}
{Garaud}, P. \& {Kulenthirarajah}, L. 2016, \apj, 821, 49

\bibitem[{{Gilmore} {et~al.}(2012){Gilmore}, {Randich}, {Asplund}, {Binney},
  {Bonifacio}, {Drew}, {Feltzing}, {Ferguson}, {Jeffries}, {Micela}, \&
  et~al.}]{Gil}
{Gilmore}, G., {Randich}, S., {Asplund}, M., {et~al.} 2012, The Messenger, 147,
  25

\bibitem[{{Gon{\c{c}}alves} {et~al.}(2020){Gon{\c{c}}alves}, {da Costa}, {de
  Almeida}, {Castro}, \& {do Nascimento}}]{goncalves20}
{Gon{\c{c}}alves}, B.~F.~O., {da Costa}, J.~S., {de Almeida}, L., {Castro}, M.,
  \& {do Nascimento}, J.~D., J. 2020, \mnras, 498, 2295

\bibitem[{{Gonzalez} {et~al.}(2009){Gonzalez}, {Zoccali}, {Monaco}, {Hill},
  {Cassisi}, {Minniti}, {Renzini}, {Barbuy}, {Ortolani}, \&
  {Gomez}}]{gonzalez09}
{Gonzalez}, O.~A., {Zoccali}, M., {Monaco}, L., {et~al.} 2009, \aap, 508, 289

\bibitem[{{Gratton} {et~al.}(2000){Gratton}, {Sneden}, {Carretta}, \&
  {Bragaglia}}]{gratton00}
{Gratton}, R.~G., {Sneden}, C., {Carretta}, E., \& {Bragaglia}, A. 2000, \aap,
  354, 169

\bibitem[{{Green} {et~al.}(2019){Green}, {Schlafly}, {Zucker}, {Speagle}, \&
  {Finkbeiner}}]{green19}
{Green}, G.~M., {Schlafly}, E., {Zucker}, C., {Speagle}, J.~S., \&
  {Finkbeiner}, D. 2019, \apj, 887, 93

\bibitem[{{Grisoni} {et~al.}(2019){Grisoni}, {Matteucci}, {Romano}, \&
  {Fu}}]{2019MNRAS.489.3539G}
{Grisoni}, V., {Matteucci}, F., {Romano}, D., \& {Fu}, X. 2019, \mnras, 489,
  3539

\bibitem[{{Guiglion} {et~al.}(2016){Guiglion}, {de Laverny}, {Recio-Blanco},
  {Worley}, {De Pascale}, {Masseron}, {Prantzos}, \& {Mikolaitis}}]{Guiglion16}
{Guiglion}, G., {de Laverny}, P., {Recio-Blanco}, A., {et~al.} 2016, \aap, 595,
  A18

\bibitem[{{Henkel} {et~al.}(2017){Henkel}, {Karakas}, \&
  {Lattanzio}}]{2017MNRAS.469.4600H}
{Henkel}, K., {Karakas}, A.~I., \& {Lattanzio}, J.~C. 2017, \mnras, 469, 4600

\bibitem[{{Hill} \& {Pasquini}(1999)}]{hill99}
{Hill}, V. \& {Pasquini}, L. 1999, \aap, 348, L21

\bibitem[{{Holanda} {et~al.}(2020){Holanda}, {Drake}, \& {Pereira}}]{holanda20}
{Holanda}, N., {Drake}, N.~A., \& {Pereira}, C.~B. 2020, \mnras, 498, 77

\bibitem[{{Iben}(1967)}]{1967ApJ...147..624I}
{Iben}, Icko, J. 1967, \apj, 147, 624

\bibitem[{{Jackson} {et~al.}(2015){Jackson}, {Jeffries}, {Lewis}, {Koposov},
  {Sacco}, {Randich}, {Gilmore}, {Asplund}, {Binney}, {Bonifacio}, {Drew},
  {Feltzing}, {Ferguson}, {Micela}, {Neguerela}, {Prusti}, {Rix}, {Vallenari},
  {Alfaro}, {Allende Prieto}, {Babusiaux}, {Bensby}, {Blomme}, {Bragaglia},
  {Flaccomio}, {Francois}, {Hambly}, {Irwin}, {Korn}, {Lanzafame}, {Pancino},
  {Recio-Blanco}, {Smiljanic}, {Van Eck}, {Walton}, {Bayo}, {Bergemann},
  {Carraro}, {Costado}, {Damiani}, {Edvardsson}, {Franciosini}, {Frasca},
  {Heiter}, {Hill}, {Hourihane}, {Jofr{\'e}}, {Lardo}, {de Laverny}, {Lind},
  {Magrini}, {Marconi}, {Martayan}, {Masseron}, {Monaco}, {Morbidelli},
  {Prisinzano}, {Sbordone}, {Sousa}, {Worley}, \& {Zaggia}}]{jackson15}
{Jackson}, R.~J., {Jeffries}, R.~D., {Lewis}, J., {et~al.} 2015, \aap, 580, A75

\bibitem[{{Jorissen} {et~al.}(2020){Jorissen}, {Van Winckel}, {Siess},
  {Escorza}, {Pourbaix}, \& {Van Eck}}]{jorissen20}
{Jorissen}, A., {Van Winckel}, H., {Siess}, L., {et~al.} 2020, \aap, 639, A7

\bibitem[{{Kaiser} {et~al.}(2021){Kaiser}, {Clemens}, {Blouin}, {Dufour},
  {Hegedus}, {Reding}, \& {B{\'e}dard}}]{kaiser21}
{Kaiser}, B.~C., {Clemens}, J.~C., {Blouin}, S., {et~al.} 2021, Science, 371,
  168

\bibitem[{{Klein} {et~al.}(2021){Klein}, {Doyle}, {Zuckerman}, {Dufour},
  {Blouin}, {Melis}, {Weinberger}, \& {Young}}]{klein21}
{Klein}, B., {Doyle}, A.~E., {Zuckerman}, B., {et~al.} 2021, arXiv e-prints,
  arXiv:2102.01834

\bibitem[{{Krolikowski} {et~al.}(2016){Krolikowski}, {Steinhauer},
  {Deliyannis}, {Twarog}, \& {Anthony-Twarog}}]{krolikowski16}
{Krolikowski}, D.~M., {Steinhauer}, A.~J., {Deliyannis}, C.~P., {Twarog},
  B.~A., \& {Anthony-Twarog}, B.~J. 2016, in American Astronomical Society
  Meeting Abstracts, Vol. 227, American Astronomical Society Meeting Abstracts
  \#227, 240.30

\bibitem[{{Kumar} \& {Reddy}(2020)}]{kumar20a}
{Kumar}, Y.~B. \& {Reddy}, B.~E. 2020, Journal of Astrophysics and Astronomy,
  41, 49

\bibitem[{{Kumar} {et~al.}(2020){Kumar}, {Reddy}, {Campbell}, {Maben}, {Zhao},
  \& {Ting}}]{kumar20b}
{Kumar}, Y.~B., {Reddy}, B.~E., {Campbell}, S.~W., {et~al.} 2020, Nature
  Astronomy, 4, 1059

\bibitem[{{Kumar} {et~al.}(2011){Kumar}, {Reddy}, \& {Lambert}}]{kumar11}
{Kumar}, Y.~B., {Reddy}, B.~E., \& {Lambert}, D.~L. 2011, \apjl, 730, L12

\bibitem[{{Lagarde} {et~al.}(2012){Lagarde}, {Decressin}, {Charbonnel},
  {Eggenberger}, {Ekstr{\"o}m}, \& {Palacios}}]{lagarde12}
{Lagarde}, N., {Decressin}, T., {Charbonnel}, C., {et~al.} 2012, \aap, 543,
  A108

\bibitem[{{Lagarde} {et~al.}(2015){Lagarde}, {Miglio}, {Eggenberger}, {Morel},
  {Montalb{\'a}n}, {Mosser}, {Rodrigues}, {Girardi}, {Rainer}, {Poretti},
  {Barban}, {Hekker}, {Kallinger}, {Valentini}, {Carrier}, {Hareter},
  {Mantegazza}, {Elsworth}, {Michel}, \& {Baglin}}]{lagarde15}
{Lagarde}, N., {Miglio}, A., {Eggenberger}, P., {et~al.} 2015, \aap, 580, A141

\bibitem[{{Lambert} {et~al.}(1980){Lambert}, {Dominy}, \&
  {Sivertsen}}]{lambert80}
{Lambert}, D.~L., {Dominy}, J.~F., \& {Sivertsen}, S. 1980, \apj, 235, 114

\bibitem[{{Lanzafame} {et~al.}(2015){Lanzafame}, {Frasca}, {Damiani},
  {Franciosini}, {Cottaar}, {Sousa}, {Tabernero}, {Klutsch}, {Spina}, {Biazzo},
  {Prisinzano}, {Sacco}, {Randich}, {Brugaletta}, {Delgado Mena}, {Adibekyan},
  {Montes}, {Bonito}, {Gameiro}, {Alcal{\'a}}, {Gonz{\'a}lez Hern{\'a}ndez},
  {Jeffries}, {Messina}, {Meyer}, {Gilmore}, {Asplund}, {Binney}, {Bonifacio},
  {Drew}, {Feltzing}, {Ferguson}, {Micela}, {Negueruela}, {Prusti}, {Rix},
  {Vallenari}, {Alfaro}, {Allende Prieto}, {Babusiaux}, {Bensby}, {Blomme},
  {Bragaglia}, {Flaccomio}, {Francois}, {Hambly}, {Irwin}, {Koposov}, {Korn},
  {Smiljanic}, {Van Eck}, {Walton}, {Bayo}, {Bergemann}, {Carraro}, {Costado},
  {Edvardsson}, {Heiter}, {Hill}, {Hourihane}, {Jackson}, {Jofr{\'e}}, {Lardo},
  {Lewis}, {Lind}, {Magrini}, {Marconi}, {Martayan}, {Masseron}, {Monaco},
  {Morbidelli}, {Sbordone}, {Worley}, \& {Zaggia}}]{lanzafame15}
{Lanzafame}, A.~C., {Frasca}, A., {Damiani}, F., {et~al.} 2015, \aap, 576, A80

\bibitem[{{Lattanzio} {et~al.}(2015){Lattanzio}, {Siess}, {Church}, {Angelou},
  {Stancliffe}, {Doherty}, {Stephen}, \& {Campbell}}]{lattanzio15}
{Lattanzio}, J.~C., {Siess}, L., {Church}, R.~P., {et~al.} 2015, \mnras, 446,
  2673

\bibitem[{{L{\`e}bre} {et~al.}(2006{\natexlab{a}}){L{\`e}bre}, {de Laverny},
  {Do Nascimento}, \& {de Medeiros}}]{2006A&A...450.1173L}
{L{\`e}bre}, A., {de Laverny}, P., {Do Nascimento}, J.~D., J., \& {de
  Medeiros}, J.~R. 2006{\natexlab{a}}, \aap, 450, 1173

\bibitem[{{L{\`e}bre} {et~al.}(2006{\natexlab{b}}){L{\`e}bre}, {de Laverny},
  {Do Nascimento}, \& {de Medeiros}}]{lebre06}
{L{\`e}bre}, A., {de Laverny}, P., {Do Nascimento}, J.~D., J., \& {de
  Medeiros}, J.~R. 2006{\natexlab{b}}, \aap, 450, 1173

\bibitem[{{Lebzelter} {et~al.}(2012){Lebzelter}, {Uttenthaler}, {Busso},
  {Schultheis}, \& {Aringer}}]{lebzelter12}
{Lebzelter}, T., {Uttenthaler}, S., {Busso}, M., {Schultheis}, M., \&
  {Aringer}, B. 2012, \aap, 538, A36

\bibitem[{{Lind} {et~al.}(2009){Lind}, {Primas}, {Charbonnel}, {Grundahl}, \&
  {Asplund}}]{lind09}
{Lind}, K., {Primas}, F., {Charbonnel}, C., {Grundahl}, F., \& {Asplund}, M.
  2009, \aap, 503, 545

\bibitem[{{Liu} {et~al.}(2016){Liu}, {Asplund}, {Yong}, {Mel{\'e}ndez},
  {Ram{\'\i}rez}, {Karakas}, {Carlos}, \& {Marino}}]{liu16}
{Liu}, F., {Asplund}, M., {Yong}, D., {et~al.} 2016, \mnras, 463, 696

\bibitem[{{Mallik} {et~al.}(2003){Mallik}, {Parthasarathy}, \&
  {Pati}}]{mallik03}
{Mallik}, S.~V., {Parthasarathy}, M., \& {Pati}, A.~K. 2003, \aap, 409, 251

\bibitem[{{Marques} {et~al.}(2013){Marques}, {Goupil}, {Lebreton}, {Talon},
  {Palacios}, {Belkacem}, {Ouazzani}, {Mosser}, {Moya}, {Morel}, {Pichon},
  {Mathis}, {Zahn}, {Turck-Chi{\`e}ze}, \& {Nghiem}}]{2013A&A...549A..74M}
{Marques}, J.~P., {Goupil}, M.~J., {Lebreton}, Y., {et~al.} 2013, \aap, 549,
  A74

\bibitem[{{Martell} {et~al.}(2020){Martell}, {Simpson}, {Balasubramaniam},
  {Buder}, {Sharma}, {Hon}, {Stello}, {Ting}, {Asplund}, {Bland -Hawthorn}, {De
  Silva}, {Freeman}, {Hayden}, {Kos}, {Lewis}, {Lind}, {Zucker}, {Zwitter},
  {Campbell}, {Cotar}, {Horner}, {Montet}, \& {Wittenmyer}}]{martell20}
{Martell}, S., {Simpson}, J., {Balasubramaniam}, A., {et~al.} 2020, arXiv
  e-prints, arXiv:2006.02106

\bibitem[{{Mathis}(2013)}]{2013LNP...865...23M}
{Mathis}, S. 2013, {Transport Processes in Stellar Interiors}, ed. M.~{Goupil},
  K.~{Belkacem}, C.~{Neiner}, F.~{Ligni{\`e}res}, \& J.~J. {Green}, Vol. 865,
  23

\bibitem[{{Mathis} {et~al.}(2018){Mathis}, {Prat}, {Amard}, {Charbonnel},
  {Palacios}, {Lagarde}, \& {Eggenberger}}]{2018A&A...620A..22M}
{Mathis}, S., {Prat}, V., {Amard}, L., {et~al.} 2018, \aap, 620, A22

\bibitem[{{Matteucci} {et~al.}(1995){Matteucci}, {D'Antona}, \&
  {Timmes}}]{Matteucci1995}
{Matteucci}, F., {D'Antona}, F., \& {Timmes}, F.~X. 1995, \aap, 303, 460

\bibitem[{{Michaud}(1986)}]{1986ApJ...302..650M}
{Michaud}, G. 1986, \apj, 302, 650

\bibitem[{{Monaco} {et~al.}(2014){Monaco}, {Boffin}, {Bonifacio}, {Villanova},
  {Carraro}, {Caffau}, {Steffen}, {Ahumada}, {Beletsky}, \&
  {Beccari}}]{monaco14}
{Monaco}, L., {Boffin}, H.~M.~J., {Bonifacio}, P., {et~al.} 2014, \aap, 564, L6

\bibitem[{{Olive}(2013)}]{2013AIPC.1560..314O}
{Olive}, K.~A. 2013, in American Institute of Physics Conference Series, Vol.
  1560, American Institute of Physics Conference Series, ed. B.~{Fleming},
  314--321

\bibitem[{{Pace} {et~al.}(2012){Pace}, {Castro}, {Mel{\'e}ndez}, {Th{\'e}ado},
  \& {do Nascimento}}]{pace12}
{Pace}, G., {Castro}, M., {Mel{\'e}ndez}, J., {Th{\'e}ado}, S., \& {do
  Nascimento}, J.~D., J. 2012, \aap, 541, A150

\bibitem[{{Palacios} {et~al.}(2001){Palacios}, {Charbonnel}, \&
  {Forestini}}]{2001A&A...375L...9P}
{Palacios}, A., {Charbonnel}, C., \& {Forestini}, M. 2001, \aap, 375, L9

\bibitem[{{Palacios} {et~al.}(2003){Palacios}, {Talon}, {Charbonnel}, \&
  {Forestini}}]{palacios03}
{Palacios}, A., {Talon}, S., {Charbonnel}, C., \& {Forestini}, M. 2003, \aap,
  399, 603

\bibitem[{{Palmerini} {et~al.}(2011){Palmerini}, {Cristallo}, {Busso}, {Abia},
  {Uttenthaler}, {Gialanella}, \& {Maiorca}}]{palmerini11}
{Palmerini}, S., {Cristallo}, S., {Busso}, M., {et~al.} 2011, \apj, 741, 26

\bibitem[{{Pancino} {et~al.}(2017){Pancino}, {Lardo}, {Altavilla}, {Marinoni},
  {Ragaini}, {Cocozza}, {Bellazzini}, {Sabbi}, {Zoccali}, {Donati}, {Heiter},
  {Koposov}, {Blomme}, {Morel}, {S{\'{\i}}mon-D{\'{\i}}az}, {Lobel},
  {Soubiran}, {Montalban}, {Valentini}, {Casey}, {Blanco-Cuaresma},
  {Jofr{\'e}}, {Worley}, {Magrini}, {Hourihane}, {Fran{\c c}ois}, {Feltzing},
  {Gilmore}, {Randich}, {Asplund}, {Bonifacio}, {Drew}, {Jeffries}, {Micela},
  {Vallenari}, {Alfaro}, {Allende Prieto}, {Babusiaux}, {Bensby}, {Bragaglia},
  {Flaccomio}, {Hambly}, {Korn}, {Lanzafame}, {Smiljanic}, {Van Eck}, {Walton},
  {Bayo}, {Carraro}, {Costado}, {Damiani}, {Edvardsson}, {Franciosini},
  {Frasca}, {Lewis}, {Monaco}, {Morbidelli}, {Prisinzano}, {Sacco}, {Sbordone},
  {Sousa}, {Zaggia}, \& {Koch}}]{pancino17}
{Pancino}, E., {Lardo}, C., {Altavilla}, G., {et~al.} 2017, \aap, 598, A5

\bibitem[{{Pasquini} {et~al.}(2004){Pasquini}, {Randich}, {Zoccali}, {Hill},
  {Charbonnel}, \& {Nordstr{\"o}m}}]{pasquini04}
{Pasquini}, L., {Randich}, S., {Zoccali}, M., {et~al.} 2004, \aap, 424, 951

\bibitem[{{Pitrou} {et~al.}(2018){Pitrou}, {Coc}, {Uzan}, \&
  {Vangioni}}]{pitrou18}
{Pitrou}, C., {Coc}, A., {Uzan}, J.-P., \& {Vangioni}, E. 2018, \physrep, 754,
  1

\bibitem[{{Prantzos}(2012)}]{2012A&A...542A..67P}
{Prantzos}, N. 2012, \aap, 542, A67

\bibitem[{{Prat} \& {Ligni{\`e}res}(2013)}]{2013A&A...551L...3P}
{Prat}, V. \& {Ligni{\`e}res}, F. 2013, \aap, 551, L3

\bibitem[{{Prat} {et~al.}(2015){Prat}, {Ligni{\`e}res}, \&
  {Lagarde}}]{2015sf2a.conf..419P}
{Prat}, V., {Ligni{\`e}res}, F., \& {Lagarde}, N. 2015, in SF2A-2015:
  Proceedings of the Annual meeting of the French Society of Astronomy and
  Astrophysics, 419--422

\bibitem[{{Privitera} {et~al.}(2016){Privitera}, {Meynet}, {Eggenberger},
  {Vidotto}, {Villaver}, \& {Bianda}}]{privitera16}
{Privitera}, G., {Meynet}, G., {Eggenberger}, P., {et~al.} 2016, \aap, 593,
  A128

\bibitem[{{Randich} {et~al.}(2013){Randich}, {Gilmore}, \& {Gaia-ESO
  Consortium}}]{randich13}
{Randich}, S., {Gilmore}, G., \& {Gaia-ESO Consortium}. 2013, The Messenger,
  154, 47

\bibitem[{{Randich} \& {Magrini}(2021)}]{2021FrASS...8....6R}
{Randich}, S. \& {Magrini}, L. 2021, Frontiers in Astronomy and Space Sciences,
  8, 6

\bibitem[{{Randich} {et~al.}(2020){Randich}, {Pasquini}, {Franciosini},
  {Magrini}, {Jackson}, {Jeffries}, {d'Orazi}, {Romano}, {Sanna},
  {Tautvai{\v{s}}ien{\.{e}}}, {Tsantaki}, {Wright}, {Gilmore}, {Bensby},
  {Bragaglia}, {Pancino}, {Smiljanic}, {Bayo}, {Carraro}, {Gonneau},
  {Hourihane}, {Morbidelli}, \& {Worley}}]{randich20}
{Randich}, S., {Pasquini}, L., {Franciosini}, E., {et~al.} 2020, \aap, 640, L1

\bibitem[{{Randich} {et~al.}(2002){Randich}, {Primas}, {Pasquini}, \&
  {Pallavicini}}]{randich02}
{Randich}, S., {Primas}, F., {Pasquini}, L., \& {Pallavicini}, R. 2002, \aap,
  387, 222

\bibitem[{{Randich} {et~al.}(2007){Randich}, {Primas}, {Pasquini}, {Sestito},
  \& {Pallavicini}}]{randich07}
{Randich}, S., {Primas}, F., {Pasquini}, L., {Sestito}, P., \& {Pallavicini},
  R. 2007, \aap, 469, 163

\bibitem[{{Randich} {et~al.}(2018){Randich}, {Tognelli}, {Jackson}, {Jeffries},
  {Degl'Innocenti}, {Pancino}, {Re Fiorentin}, {Spagna}, {Sacco}, {Bragaglia},
  {Magrini}, {Prada Moroni}, {Alfaro}, {Franciosini}, {Morbidelli},
  {Roccatagliata}, {Bouy}, {Bravi}, {Jim{\'e}nez-Esteban}, {Jordi}, {Zari},
  {Tautvai{\v{s}}iene}, {Drazdauskas}, {Mikolaitis}, {Gilmore}, {Feltzing},
  {Vallenari}, {Bensby}, {Koposov}, {Korn}, {Lanzafame}, {Smiljanic}, {Bayo},
  {Carraro}, {Costado}, {Heiter}, {Hourihane}, {Jofr{\'e}}, {Lewis}, {Monaco},
  {Prisinzano}, {Sbordone}, {Sousa}, {Worley}, \& {Zaggia}}]{randich18}
{Randich}, S., {Tognelli}, E., {Jackson}, R., {et~al.} 2018, \aap, 612, A99

\bibitem[{{Richard} \& {Zahn}(1999)}]{1999A&A...347..734R}
{Richard}, D. \& {Zahn}, J.-P. 1999, \aap, 347, 734

\bibitem[{{Richard} {et~al.}(1996){Richard}, {Vauclair}, {Charbonnel}, \&
  {Dziembowski}}]{Richard1996}
{Richard}, O., {Vauclair}, S., {Charbonnel}, C., \& {Dziembowski}, W.~A. 1996,
  \aap, 312, 1000

\bibitem[{{Roccatagliata} {et~al.}(2018){Roccatagliata}, {Sacco},
  {Franciosini}, \& {Randich}}]{roccatagliata18}
{Roccatagliata}, V., {Sacco}, G.~G., {Franciosini}, E., \& {Randich}, S. 2018,
  \aap, 617, L4

\bibitem[{{Romano} {et~al.}(2001){Romano}, {Matteucci}, {Ventura}, \&
  {D'Antona}}]{Romano2001}
{Romano}, D., {Matteucci}, F., {Ventura}, P., \& {D'Antona}, F. 2001, \aap,
  374, 646

\bibitem[{{Sacco} {et~al.}(2014){Sacco}, {Morbidelli}, {Franciosini},
  {Maiorca}, {Randich}, {Modigliani}, {Gilmore}, {Asplund}, {Binney},
  {Bonifacio}, {Drew}, {Feltzing}, {Ferguson}, {Jeffries}, {Micela},
  {Negueruela}, {Prusti}, {Rix}, {Vallenari}, {Alfaro}, {Allende Prieto},
  {Babusiaux}, {Bensby}, {Blomme}, {Bragaglia}, {Flaccomio}, {Francois},
  {Hambly}, {Irwin}, {Koposov}, {Korn}, {Lanzafame}, {Pancino}, {Recio-Blanco},
  {Smiljanic}, {Van Eck}, {Walton}, {Bergemann}, {Costado}, {de Laverny},
  {Heiter}, {Hill}, {Hourihane}, {Jackson}, {Jofre}, {Lewis}, {Lind}, {Lardo},
  {Magrini}, {Masseron}, {Prisinzano}, \& {Worley}}]{sacco14}
{Sacco}, G.~G., {Morbidelli}, L., {Franciosini}, E., {et~al.} 2014, \aap, 565,
  A113

\bibitem[{{Sackmann} \& {Boothroyd}(1999)}]{1999ApJ...510..217S}
{Sackmann}, I.~J. \& {Boothroyd}, A.~I. 1999, \apj, 510, 217

\bibitem[{{Schlegel} {et~al.}(1998){Schlegel}, {Finkbeiner}, \&
  {Davis}}]{schlegen98}
{Schlegel}, D.~J., {Finkbeiner}, D.~P., \& {Davis}, M. 1998, \apj, 500, 525

\bibitem[{{Schramm} {et~al.}(1990){Schramm}, {Steigman}, \&
  {Dearborn}}]{1990ApJ...359L..55S}
{Schramm}, D.~N., {Steigman}, G., \& {Dearborn}, D. S.~P. 1990, \apjl, 359, L55

\bibitem[{{Sengupta} \& {Garaud}(2018{\natexlab{a}})}]{2018ApJ...862..136S}
{Sengupta}, S. \& {Garaud}, P. 2018{\natexlab{a}}, \apj, 862, 136

\bibitem[{{Sengupta} \& {Garaud}(2018{\natexlab{b}})}]{Sengupta18}
{Sengupta}, S. \& {Garaud}, P. 2018{\natexlab{b}}, \apj, 862, 136

\bibitem[{{Sestito} \& {Randich}(2005)}]{SR05}
{Sestito}, P. \& {Randich}, S. 2005, \aap, 442, 615

\bibitem[{{Siess} \& {Livio}(1999)}]{1999MNRAS.308.1133S}
{Siess}, L. \& {Livio}, M. 1999, \mnras, 308, 1133

\bibitem[{{Singh} {et~al.}(2021){Singh}, {Reddy}, {Campbell}, {Bharat Kumar},
  \& {Vrard}}]{singh21}
{Singh}, R., {Reddy}, B.~E., {Campbell}, S.~W., {Bharat Kumar}, Y., \& {Vrard},
  M. 2021, arXiv e-prints, arXiv:2104.12070

\bibitem[{{Skrutskie} {et~al.}(2006){Skrutskie}, {Cutri}, {Stiening},
  {Weinberg}, {Schneider}, {Carpenter}, {Beichman}, {Capps}, {Chester},
  {Elias}, {Huchra}, {Liebert}, {Lonsdale}, {Monet}, {Price}, {Seitzer},
  {Jarrett}, {Kirkpatrick}, {Gizis}, {Howard}, {Evans}, {Fowler}, {Fullmer},
  {Hurt}, {Light}, {Kopan}, {Marsh}, {McCallon}, {Tam}, {Van Dyk}, \&
  {Wheelock}}]{2mass}
{Skrutskie}, M.~F., {Cutri}, R.~M., {Stiening}, R., {et~al.} 2006, \aj, 131,
  1163

\bibitem[{{Smiljanic}(2020)}]{2020MmSAI..91..142S}
{Smiljanic}, R. 2020, \memsai, 91, 142

\bibitem[{{Smiljanic} {et~al.}(2018){Smiljanic}, {Franciosini}, {Bragaglia},
  {Tautvai{\v{s}}ien{\.{e}}}, {Fu}, {Pancino}, {Adibekyan}, {Sousa}, {Randich},
  {Montalb{\'a}n}, {Pasquini}, {Magrini}, {Drazdauskas}, {Garc{\'\i}a},
  {Mathur}, {Mosser}, {R{\'e}gulo}, {de Assis Peralta}, {Hekker}, {Feuillet},
  {Valentini}, {Morel}, {Martell}, {Gilmore}, {Feltzing}, {Vallenari},
  {Bensby}, {Korn}, {Lanzafame}, {Recio-Blanco}, {Bayo}, {Carraro}, {Costado},
  {Frasca}, {Jofr{\'e}}, {Lardo}, {de Laverny}, {Lind}, {Masseron}, {Monaco},
  {Morbidelli}, {Prisinzano}, {Sbordone}, \& {Zaggia}}]{smi18}
{Smiljanic}, R., {Franciosini}, E., {Bragaglia}, A., {et~al.} 2018, \aap, 617,
  A4

\bibitem[{{Smiljanic} {et~al.}(2009){Smiljanic}, {Gauderon}, {North}, {Barbuy},
  {Charbonnel}, \& {Mowlavi}}]{smi09}
{Smiljanic}, R., {Gauderon}, R., {North}, P., {et~al.} 2009, \aap, 502, 267

\bibitem[{{Smiljanic} {et~al.}(2014){Smiljanic}, {Korn}, {Bergemann}, {Frasca},
  {Magrini}, {Masseron}, {Pancino}, {Ruchti}, {San Roman}, {Sbordone}, {Sousa},
  {Tabernero}, {Tautvai{\v s}ien{\.e}}, {Valentini}, {Weber}, {Worley},
  {Adibekyan}, {Allende Prieto}, {Barisevi{\v c}ius}, {Biazzo},
  {Blanco-Cuaresma}, {Bonifacio}, {Bragaglia}, {Caffau}, {Cantat-Gaudin},
  {Chorniy}, {de Laverny}, {Delgado-Mena}, {Donati}, {Duffau}, {Franciosini},
  {Friel}, {Geisler}, {Gonz{\'a}lez Hern{\'a}ndez}, {Gruyters}, {Guiglion},
  {Hansen}, {Heiter}, {Hill}, {Jacobson}, {Jofre}, {J{\"o}nsson}, {Lanzafame},
  {Lardo}, {Ludwig}, {Maiorca}, {Mikolaitis}, {Montes}, {Morel}, {Mucciarelli},
  {Mu{\~n}oz}, {Nordlander}, {Pasquini}, {Puzeras}, {Recio-Blanco}, {Ryde},
  {Sacco}, {Santos}, {Serenelli}, {Sordo}, {Soubiran}, {Spina}, {Steffen},
  {Vallenari}, {Van Eck}, {Villanova}, {Gilmore}, {Randich}, {Asplund},
  {Binney}, {Drew}, {Feltzing}, {Ferguson}, {Jeffries}, {Micela}, {Negueruela},
  {Prusti}, {Rix}, {Alfaro}, {Babusiaux}, {Bensby}, {Blomme}, {Flaccomio},
  {Fran{\c c}ois}, {Irwin}, {Koposov}, {Walton}, {Bayo}, {Carraro}, {Costado},
  {Damiani}, {Edvardsson}, {Hourihane}, {Jackson}, {Lewis}, {Lind}, {Marconi},
  {Martayan}, {Monaco}, {Morbidelli}, {Prisinzano}, \& {Zaggia}}]{smi14}
{Smiljanic}, R., {Korn}, A.~J., {Bergemann}, M., {et~al.} 2014, \aap, 570, A122

\bibitem[{{Smiljanic} {et~al.}(2010){Smiljanic}, {Pasquini}, {Charbonnel}, \&
  {Lagarde}}]{smi10}
{Smiljanic}, R., {Pasquini}, L., {Charbonnel}, C., \& {Lagarde}, N. 2010, \aap,
  510, A50

\bibitem[{{Somers} \& {Pinsonneault}(2016)}]{2016ApJ...829...32S}
{Somers}, G. \& {Pinsonneault}, M.~H. 2016, \apj, 829, 32

\bibitem[{{Talon} \& {Charbonnel}(2005)}]{CT05}
{Talon}, S. \& {Charbonnel}, C. 2005, \aap, 440, 981

\bibitem[{{Talon} \& {Charbonnel}(2010)}]{2010IAUS..268..365T}
{Talon}, S. \& {Charbonnel}, C. 2010, in IAU Symposium, Vol. 268, Light
  Elements in the Universe, ed. C.~{Charbonnel}, M.~{Tosi}, F.~{Primas}, \&
  C.~{Chiappini}, 365--374

\bibitem[{{Talon} \& {Zahn}(1997)}]{TZ1997}
{Talon}, S. \& {Zahn}, J.~P. 1997, \aap, 317, 749

\bibitem[{{Tognelli} {et~al.}(2020){Tognelli}, {Prada Moroni},
  {Degl'Innocenti}, {Salaris}, \& {Cassisi}}]{tognelli20}
{Tognelli}, E., {Prada Moroni}, P.~G., {Degl'Innocenti}, S., {Salaris}, M., \&
  {Cassisi}, S. 2020, \aap, 638, A81

\bibitem[{{Travaglio} {et~al.}(2001){Travaglio}, {Randich}, {Galli},
  {Lattanzio}, {Elliott}, {Forestini}, \& {Ferrini}}]{Travaglio2001}
{Travaglio}, C., {Randich}, S., {Galli}, D., {et~al.} 2001, \apj, 559, 909

\bibitem[{{Valle} {et~al.}(2014){Valle}, {Dell'Omodarme}, {Prada Moroni}, \&
  {Degl'Innocenti}}]{valle14}
{Valle}, G., {Dell'Omodarme}, M., {Prada Moroni}, P.~G., \& {Degl'Innocenti},
  S. 2014, \aap, 561, A125

\bibitem[{{Wallerstein} {et~al.}(1994){Wallerstein}, {Bohm-Vitense}, {Vanture},
  \& {Gonzalez}}]{walle94}
{Wallerstein}, G., {Bohm-Vitense}, E., {Vanture}, A.~D., \& {Gonzalez}, G.
  1994, \aj, 107, 2211

\bibitem[{{Wang} {et~al.}(2021){Wang}, {Nordlander}, {Asplund}, {Amarsi},
  {Lind}, \& {Zhou}}]{wang21}
{Wang}, E.~X., {Nordlander}, T., {Asplund}, M., {et~al.} 2021, \mnras, 500,
  2159

\bibitem[{{Yan} {et~al.}(2021){Yan}, {Zhou}, {Zhang}, {Li}, {Gao}, {Shi},
  {Zhao}, {Aoki}, {Matsuno}, {Li}, {Xu}, {Li}, {Wu}, {Jin}, {Mosser}, {Bi},
  {Fu}, {Pan}, {Suda}, {Liu}, {Zhao}, \& {Liang}}]{yan21}
{Yan}, H.-L., {Zhou}, Y.-T., {Zhang}, X., {et~al.} 2021, Nature Astronomy, 5,
  86

\bibitem[{{Zahn}(1992)}]{JPZ1992}
{Zahn}, J.~P. 1992, \aap, 265, 115

\bibitem[{{Zhang}(2012)}]{zhang12}
{Zhang}, Q.~S. 2012, \mnras, 427, 1441

\end{thebibliography}

\begin{appendix}
\section{}
\begin{table*}[h]
\caption{Parameters of our sample open clusters from  Gaia-ESO {\sc idr6}}
\begin{center}
\tiny
\begin{tabular}{llrrrrrll}
\hline
\hline
id & Cluster &  Age & Distance    &  R$_{\rm GC}$ & RV           & [Fe/H] & MSTO & Parsec (Age$_{\rm iso}$, Z$_{\rm iso}$) \\
   &         &  (Myr) & (pc)     & (kpc)      & (km s$^{-1}$)  & (dex)  &  ($M_{\odot}$) & (Gyr, Z) \\
\hline
1  & NGC6067    & 130 &   1880  & 6.8 & -39.4$\pm$0.2 &  0.03$\pm$0.16 & 4.5 &(0.13, 0.0145)  \\
2  & NGC6709    & 190 &   1040  & 7.6 & -11$\pm$2 & -0.03$\pm$0.03  & 3.8 &(0.2, 0.013) \\
3 & Rup7        & 230 &   5850  & 13.1&  77$\pm$1 & -0.24$\pm$0.04  & 3.4 &(0.24, 0.007) \\
4  & NGC6192    & 240 &   1740  & 6.7 & -8.1$\pm$0.7  & -0.08$\pm$0.08 & 3.5 &(0.24, 0.011) \\
5  & NGC6259    & 270 &   2310  & 6.2 & -34$\pm$1 &  0.17$\pm$0.06  & 3.5 &(0.26, 0.019) \\
6  & Berkeley30       & 295 &   5380  & 13.3&  47$\pm$2 & -0.08$\pm$0.13  & 3.2 &(0.3, 0.009)  \\
7  & NGC6705    & 310 &   2200  & 6.5 &  34$\pm$1 &  0.02$\pm$0.05  & 3.3 &(0.3, 0.015) \\
8 & NGC4815    & 370 &   3295  & 7.1 & -29.9$\pm$0.4 &  0.04$\pm$0.16  & 3.0 &(0.38, 0.015) \\
9  & NGC3532   & 400 &    500  & 8.2 &   5$\pm$1 & -0.01$\pm$0.06  & 2.9 &(0.4, 0.013) \\
10  & NGC6281   & 510 &    540  & 7.8 &  -6$\pm$1 & -0.08$\pm$0.16  & 2.6 &(0.52, 0.011) \\
11 & NGC2324    & 540 &   4210  & 12.1&  41$\pm$3 & -0.17$\pm$0.02  & 2.5 &(0.54, 0.009) \\
12 & Pismis18   & 575  &  2860  & 6.9 & -27.8$\pm$0.7 &0.14$\pm$0.04    & 2.6 &(0.58, 0.019) \\
13 & NGC6802    & 660 &   2750  & 7.1 &  11.6$\pm$0.8 &  0.14$\pm$0.04  & 2.5 &(0.66, 0.019) \\
14 & NGC6633    & 690 &    420  & 8.0 & -29$\pm$1 & -0.06$\pm$0.11  & 2.3 &(0.7, 0.011) \\
15 & Trumpler23 & 710 &   2590  & 6.3 & -62$\pm$1 &  0.21$\pm$0.03  & 2.5 &(0.7, 0.021) \\
16 & Pismis15   & 870 &   2560  & 8.6 &  34.8$\pm$0.5 &  0.04$\pm$0.04 & 2.2 &(0.88, 0.015) \\
17 & NGC3960    & 870 &   2345  & 7.7 & -22$\pm$1 & -0.06$\pm$0.16  & 2.2& (0.88, 0.011)  \\
18 & NGC5822    & 910 &    850  & 7.7 & -28$\pm$2 &  0.01$\pm$0.02  & 2.2 &(0.92, 0.015) \\
19 & NGC2660    & 930  &  2790  & 9.0 &  22.0$\pm$0.3 & -0.05$\pm$0.04  & 2.1 &(0.94, 0.011) \\
20 & Melotte71  & 980  &  2140  &9.9  & 50$\pm$1  & -0.10$\pm$0.03   &2.1 &(0.98, 0.011) \\
21 & NGC2355    &1000 &   1940  & 10.1&  36.2$\pm$0.5 & -0.09$\pm$0.03  & 2.1 &(1.0, 0.011) \\
22 & NGC2477    &1120 &   1440  & 8.9 &  8$\pm$1  &  0.14$\pm$0.04  & 2.1 &(1.12, 0.019) \\
23 & Berkeley81       & 1150&   3310  & 5.9 &  47.9$\pm$0.6 &  0.22$\pm$0.06  & 2.1 &(1.15, 0.023) \\
24 & NGC6583    & 1200 &  2050  & 6.3 &  -1.9$\pm$0.9 &  0.22$\pm$0.01  & 2.0 &(1.2, 0.023) \\
25 & NGC6005    & 1260 &  2380  & 6.5 & -24$\pm$1 &0.21$\pm$0.04    & 2.0 &(1.25, 0.023) \\
26 & Berkeley73       & 1410 &  6160  &13.8 &  97.0$\pm$0.3 & -0.26$\pm$0.04  & 1.7 &(1.42, 0.007) \\
27 & Berkeley44       & 1440 &  2860  &7.0  & -9.1$\pm$0.5  &  0.22$\pm$0.10  & 1.9 &(1.45, 0.023) \\
28 & NGC4337    &1440 &   2450  & 7.5 & -18.9$\pm$0.2 &  0.25$\pm$0.02  & 1.9 &(1.45, 0.023) \\
29 & NGC2158    & 1550&   4300  & 12.6&  27$\pm$2 & -0.16$\pm$0.05  & 1.7 &(1.54, 0.009) \\
30 & Tom2       & 1620 &  9320  & 15.6& 122$\pm$1 &-0.24$\pm$0.08   & 1.7 &(1.62, 0.007) \\
31 & NGC2506$^a$    & 1660 &  3190  & 10.6&  84.1 & -0.34  & 1.6 &(1.66, 0.007) \\
32 & Rup134     & 1660 &  2250  & 6.1 & -41.1$\pm$0.6 & 0.27$\pm$0.04   & 1.8 &(1.65, 0.023) \\
33 & Berkeley75       & 1700 &  8300  & 14.7& 124$\pm$1 & -0.35$\pm$0.05  & 1.6 &(1.7, 0.005) \\
34 & NGC2420    & 1740 &  2590  & 10.7&  74.1$\pm$0.8 & -0.16$\pm$0.07  & 1.7 &(1.7, 0.005) \\
35 & Col110     & 1820 &  2180  &10.3 &  37.2$\pm$0.3 & -0.10$\pm$0.05 & 1.7 &(1.82, 0.011) \\
36 & NGC2141    & 1860 &  5180  & 13.3& 26.0$\pm$0.8  & -0.06$\pm$0.07  & 1.6 &(1.86, 0.11) \\
37 & Trumpler20 & 1860 &  3390  &7.2  & -40.2$\pm$0.9 &0.13$\pm$0.04    & 1.5 &(1.86, 0.019) \\
38 & Berkeley21       & 2140 &  6420  &14.7 & 0.9$\pm$0.9   & -0.21$\pm$0.04  & 1.5 &(2.14, 0.009)\\
39 & NGC2425    & 2400 &  3580  & 10.9& 103.5$\pm$0.7 & -0.17$\pm$0.18  & 1.5 &(2.4, 0.009) \\
40 & Berkeley22       & 2450 &  6220  & 14.3& 95$\pm$2  &-0.28$\pm$0.08    & 1.4& (2.46, 0.007) \\
41 & Berkeley25       & 2450 &  6780  & 13.8 & 136$\pm$1 & -0.30$\pm$0.12  &1.4 &(2.46, 0.007) \\
42 & Cz24       & 2700 &  3980  &12.3 & 22.1$\pm$0.5  & -0.12$\pm$0.04   & 1.4 &(2.7, 0.011) \\
43 & Berkeley31       & 2820 &  7180  & 15.1& 56.9$\pm$0.8  & -0.31$\pm$0.04   &1.3 &(2.82, 0.007) \\
44 & Cz30       & 2880 &  6650  &13.8 & 81.5$\pm$0.3  & -0.32$\pm$0.01   &1.3 &(2.88, 0.007) \\
45 & Berkeley29       & 3090 &  12600  & 20.6& 25.8$\pm$0.3  & -0.39$\pm$0.11   &1.3 &(3.0, 0.005) \\
46 & NGC6253    & 3230 & 1650 & 6.9 &-29.2$\pm$0.9  &0.33$\pm$0.10   &1.5 &(3.2, 0.029) \\
47 &  Haf10     & 3800 &  3410  &10.8 & 87.8$\pm$0.9  & -0.11$\pm$0.03   &1.2 &(3.8, 0.011) \\
48 & Trumpler5  & 4260 & 3050 &11.2 & 51$\pm$1  & -0.35$\pm$0.04  &1.1 &(4.2, 0.007)\\
49 & M67        & 4260 &  900  & 8.7 &34.1$\pm$0.6   & -0.02$\pm$0.09   &1.2 &(4.2, 0.013) \\
50 & NGC2243    & 4360 & 3720 &10.6 & 59.8$\pm$0.3  & -0.44$\pm$0.09  &1.1 &(4.4, 0.005) \\
51 & ESO92\_05 & 4470 & 12440 &12.8 &61.4$\pm$0.4   & -0.29$\pm$0.07  &1.1 &(4.4, 0.007) \\
52 & Berkeley20       & 4790 & 8730 & 16.3 & 78.9$\pm$0.3  &    -0.37$\pm$0.07  &1.1 &(4.8, 0.005) \\
53 & Berkeley32       & 4900 & 3070 &11.1 & 105.9$\pm$0.6 & -0.28$\pm$0.08  &1.1 &(4.8, 0.009) \\
54 & Berkeley39       & 5620 &3970 & 11.5 & 58.6$\pm$0.8  &    -0.14$\pm$0.05   &1.1 &(5.6, 0.009) \\
55 & NGC6791    & 6310 & 4230 &7.9  &-48$\pm$1  & 0.32$\pm$0.18*    &1.1 &(6.4, 0.019) \\
56 & Col261     & 6310 & 2850  & 7.3 &-25$\pm$1  &  -0.05$\pm$0.07    &1.1 &(6.4, 0.019) \\
57 & Berkeley36       & 6760 & 4360 &11.7 &   62.6$\pm$0.9  & -0.22$\pm$0.11  & 1.0 &(6.8, 0.009) \\
 \hline
\end{tabular}
\tablefoot{$^a$ only one star}
\label{tab:clusters}
\end{center}
\end{table*}

\begin{table*}[h]
\caption{Sample of selected member stars in open clusters. The full table is available online at CDS.}
\begin{center}
\tiny{
\begin{tabular}{lllllllllll}
\hline
\hline
id & Cluster &  T$_{\rm eff}$ & log~g   & [Fe/H] & A(Li) & UL$_{\rm A(Li)} ^a$ & log(L/L$_{\odot}$)   & $\gamma$  & Mass (MSTO) \\
   &         &  (K)           &         &  (dex) &  (dex)&                    &                   & & (M$_{\odot}$)   \\
\hline
  05323677+0011048  & Br20    &     4850$\pm$30 &    2.70$\pm$0.05  &   -0.32$\pm$0.06   & -0.02     &   1        &        1.64$\pm$0.04    &--&        1.1       \\
  05323896+0011203  & Br20    &     4380$\pm$30 &    1.81$\pm$0.06  &   -0.43$\pm$0.06   & -0.63      &  1        &            2.13$\pm$0.03    &--&         1.1      \\               
  05512981+2143071  & Br21    &     6740$\pm$60 &    4.3$\pm$0.2  &   -0.37$\pm$0.05   & 2.30$\pm$0.25 &   0    &            1.29$\pm$0.06    &--&          1.5      \\ 
  05515964+2144121  & Br21    &     6240$\pm$80 &    4.1$\pm$0.2  &   -0.28$\pm$0.08   & 2.41       &  1        &            0.8$\pm$0.1    &--&         1.5      \\
\hline
\end{tabular}
}
\tablefoot{$^a$ upper limits are indicated with 1, detections with 0}
\label{table:cluster:sample}
\end{center}
\end{table*}

\begin{table*}[h]
\caption{Sample of selected stars in the MW fields. The full table is available online at CDS.}
\begin{center}
\tiny{
\begin{tabular}{llllllllll}
\hline
\hline
id & GES\_FLD &  T$_{\rm eff}$ & log~g   & [Fe/H] & A(Li) & UL$_{\rm A(Li)} ^a$ & log(L/L$_{\odot}$)  & $\gamma$  & Mass  \\
   &         &  (K)           &         &  (dex) &  (dex)&                    &                    &       & (M$_{\odot}$)   \\
\hline
 00000009-5455467  & GES\_MW\_00\_01 &   6060$\pm$30 & 3.94$\pm$0.05 & -0.55$\pm$0.05 &  2.34$\pm$0.03 &   0      &    0.25$\pm$0.02 & --&      0.8$\pm$0.1\\   
 00000302-6002570  & GES\_MW\_00\_01 &  5780$\pm$30 & 4.04$\pm$0.05 & -0.31$\pm$0.04 &  2.01$\pm$0.03 &   0    &      0.37$\pm$0.01 & -- &      0.9$\pm$0.1\\ 
\hline
\end{tabular}
}
\tablefoot{$^a$ upper limits are indicated with 1, detections with 0}
\label{table:field:sample}
\end{center}
\end{table*}

\end{appendix}

\end{document}